\documentclass[11pt]{article}

\usepackage[margin=1in]{geometry}
\usepackage{amsmath,amssymb,amsfonts}
\usepackage{graphicx}
\usepackage{physics}
\usepackage{bm}
\usepackage{mathtools}
\usepackage{subfigure}
\usepackage{float}
\usepackage{slashed}
\usepackage{tikz}
\usepackage{cancel}
\usepackage{hyperref}
\hypersetup{
  colorlinks=true,
  linkcolor=blue,
  citecolor=blue,
  urlcolor=blue
}
\usepackage{tikz}
\usetikzlibrary{arrows.meta,decorations.pathmorphing,calc}
\usepackage{booktabs}
\usepackage{array}
\usepackage{pgfplots}
\pgfplotsset{compat=1.18}
\usepackage{xcolor}
\numberwithin{equation}{section}

\numberwithin{equation}{section}

\begin{document}

\vspace*{2.5cm}
\begin{center}
{\LARGE {Capacity of Entanglement and Replica Backreaction in RST Gravity} \\ \vspace*{1cm}}

\end{center}

\begin{center}
Ra\'ul Arias$^{\dagger}$\footnote{\texttt{rarias@fisica.unlp.edu.ar}},
Daniel Fondevila$^{\dagger}$\footnote{\texttt{dfondevila@iflp.unlp.edu.ar}}
\end{center}

\begin{center}
{\footnotesize
\vspace{0.4cm}
$^\dagger$ Instituto de F\'isica La Plata -- CONICET and
Departamento de F\'isica, Universidad Nacional de La Plata, C.C.~67, 1900 La Plata, Argentina
}
\end{center}

\vspace*{0.5cm}

\vspace*{1.5cm}
\begin{abstract}
\noindent
We compute the capacity of entanglement in two-dimensional dilaton gravity in a setting where Hawking radiation, backreaction, and islands can be treated analytically. Our focus is the eternal black hole of the Russo–Susskind–Thorlacius model coupled to N conformal matter fields. Unlike previous gravitational computations, which were mostly carried out in JT gravity, the RST model forces one to deal with a genuinely dynamical conformal factor and with the global constraints of the replica construction. The main technical step is therefore to solve the replica deformation on the orbifold globally at first order near $n=1$, including the homogeneous sector fixed by single-valuedness and by the requirement of a fixed microcanonical state. For a single interval we obtain a time-independent generalized capacity, parallel to the generalized entropy. For two intervals, even in the late-time factorization regime, the global solution generates an interaction term between replica fixed points; after Lorentzian continuation this produces a time-dependent capacity on the two-QES saddle, despite the corresponding entropy plateau. We discuss the regime of validity of the resulting expressions and explain how the large size of the two-QES capacity implies a highly non-uniform saddle competition near $n=1$, providing a concrete mechanism for sharp features of the capacity at the Page transition.

\end{abstract}

\newpage

\setcounter{tocdepth}{2}
\tableofcontents

\newpage

\section{Introduction}

The last few years have seen a rather sharp change in how we think about fine--grained entropy in
gravitational systems.  The Bekenstein--Hawking formula \cite{Bekenstein1973,Hawking1975} and the
expectation of unitary evaporation \cite{Page1993} have been with us for decades, but the practical
tool to compute the Page curve semiclassically only became available recently.  The key input is the
replica representation of $\Tr\rho_A^n$ and the observation that the gravitational path integral at
integer $n$ admits additional saddles that were previously ignored in the semiclassical expansion. In AdS
settings, these saddles appear as replica wormholes and reproduce a unitary Page curve for the
radiation entropy \cite{Penington2019,AEMM2019,AMMZ2019,AHMST2019}.  The same logic can be
phrased in the language of quantum extremal surfaces (QES) \cite{EngelhardtWall2014},
where the generalized entropy $S_{\rm gen}=A/4G_{\rm eff}+S_{\rm vN}$ is extremized over candidate
surfaces (or points, in two dimensions).  From a purely holographic perspective, this is the quantum
extension of the RT/HRT prescriptions \cite{RyuTakayanagi2006,HRT2007}, understood and derived from replicas and conical geometries constructions \cite{LewkowyczMaldacena2013} for Rényi entropies \cite{Dong2016}.  At the first quantum order, the bulk--entanglement
correction to RT is captured by the FLM formula \cite{FaulknerLewkowyczMaldacena2013}, and the
relation between boundary and bulk modular data is encoded in JLMS \cite{JLMS2016}.  These
developments make it hard to avoid the conclusion that \emph{replicas are part of the dynamics}:
nontrivial gravitational physics appear from the analytic continuation of $n\to1$.

It is natural to ask how much information is actually contained in the Page curve
itself.  The von Neumann entropy is only the first cumulant of the entanglement spectrum.  A more
refined probe is the \emph{capacity of entanglement}, defined as the variance of the modular
Hamiltonian (equivalently, the second cumulant of the entanglement spectrum).  In replica language,
\begin{equation}
C(A)=\lim_{n\to1} n^2\,\partial_n^2 \log \Tr\rho_A^n,
\end{equation}
so $C$ is sensitive to the replica path integral in a way that $S$ is not.  The capacity has been
studied in QFT and holography from various viewpoints \cite{deBoerJarvelaKeskiVakkuri2019, Arias2023},
including dynamical settings where it carries information beyond entanglement entropy itself
\cite{Nandy2021}.  In gravitational systems, Kawabata, Nishioka, Okuyama and Watanabe proposed a
replica--wormhole prescription for the capacity and emphasized that it may develop a sharp feature
(jump or peak) at the Page transition even when $S_{\rm gen}$ is continuous \cite{KawabataEtAl2102,KawabataEtAl2105}.
This is, in a precise sense, the simplest quantity for which one is forced to confront the detailed
$n$--dependence of the saddle beyond what is needed to compute the entropy.

Most explicit gravitational computations of the capacity so far have been carried out in AdS$_2$/JT
setups \cite{KawabataEtAl2102,KawabataEtAl2105, Yu2025}, where the replica geometry details were managed only in the high temperature limit due to the welding problem. \color{black} By contrast, our goal here is to
compute the capacity in an \emph{asymptotically flat} two--dimensional model where the Hawking
radiation is represented by a large--$N$ conformal sector and the semiclassical backreaction is
exactly tractable: the Russo--Susskind--Thorlacius (RST) model \cite{CGHS1992,RST1992,Polyakov1981}.
The RST theory is the quantum--corrected completion of CGHS \cite{CGHS1992}, where the Polyakov
action \cite{Polyakov1981} accounts for the conformal anomaly and the additional RST counterterm
restores solvability \cite{RST1992}.  The model has become a canonical arena for islands in
asymptotically flat evaporation, and in particular Hartman, Shaghoulian and Strominger used it to
give a clean derivation of the island Page curve in flat space \cite{HartmanShaghoulianStrominger2020},
complemented by related analyses in the same theory \cite{GautasonEtAl2020, Anegawa2020}.  What is still missing,
and what becomes unavoidable for the capacity, is an explicit control of the \emph{global} replica
deformation at order $(n-1)$.

This point deserves emphasis.  Near a $\mathbb{Z}_n$ fixed point in the replica geometry, one can always introduce a local
uniformizing coordinate in which the cover is smooth and the orbifold looks like a cone. For the entropy, however, one typically does not need the full \emph{off-shell} replica backreaction:
in the replica derivation of generalized entropy the area term already enters at order \((n-1)\),
so \(S_{\rm gen}\) can be obtained by evaluating the appropriate local term on the \(n=1\) background,
once the state is fixed by the asymptotic charges and the QES location is determined.  The capacity
is different.  The second derivative in $n$ forces one to keep track of the $n$--dependence of the
conformal factor and of the fields variables of RST.  In
practice, this means that one must solve the linearized replica equations globally and fix the
homogeneous modes by appropriate global constraints (single--valuedness, asymptotic frame fixing,
and fixed--state conditions).  In this paper we do precisely this in the eternal RST black hole, in a
way that makes explicit the difference between local and global data. The technical core of the paper is therefore a global solution of the replica equations on the
orbifold, at first order in $(n-1)$, both for one fixed point (one-QES) and for two fixed points
(two-QES).  

For a single interval we show that a simple $U(1)$--symmetric ansatz near the replica fixed point,
combined with the correct identification of the thermal circle on the orbifold and the requirement
that the underlying black hole state is kept fixed, matches uniquely onto the global solution and
fixes the remaining homogeneous mode. A useful point is that ``keeping the state fixed'' does not
mean that every constant appearing in the orbifold description is independent of the replica index:
the orbifold has a shorter Euclidean time circle, so the same physical state is represented with a
different normalization of the constant term in the local solution. This $n$--dependence is
therefore kinematical rather than a change of state. With these identifications, the one--interval
generalized capacity is time independent, in parallel with the generalized entropy.

The two--interval configuration is more interesting.  In the late--time factorization regime, the
matter source in the replica constraint reduces to a sum of two one--interval sources, but the
\emph{global} solution is still a single problem with one set of homogeneous modes.  Evaluating the
capacity on this two--fixed--point orbifold produces an ``interaction'' term that depends on the
invariant separation $\Delta=|z_1-z_2|$ between the fixed points.  After continuing to Lorentzian
signature with the appropriate two--sided identification, $\Delta$ becomes time dependent, and so does the two--QES capacity.  This is the main qualitative difference with respect to the entropy in the same configuration: while the generalized entropy on the island saddle is time independent in the factorization regime, the capacity retains a nontrivial dependence on $t$ through $\Delta(t)$.  In addition, the large size of the two--QES capacity in Lorentzian kinematics implies that saddle competition in the $(n,t)$ plane can be highly non--uniform near
$n=1$, providing a concrete mechanism for sharp features of the capacity at the von Neumann Page
transition along the lines of \cite{KawabataEtAl2105}.  It is important, however, to keep track of
the regime of validity: our gravitational computation is perturbative in $(n-1)$ and therefore
controls the phase boundary only in a neighborhood of $n=1$.

The paper is organized as follows.  In Sec. \ref{Sec: reviewRST} we review the RST model and the eternal black hole,
and we summarize the generalized entropy for one and two intervals in the kinematics relevant to
the Hartman--Shaghoulian--Strominger configuration \cite{HartmanShaghoulianStrominger2020}.
In Sec. \ref{sec:replica-eqs-CoE} we derive the replica equations on the orbifold and solve them globally at order $(n-1)$,
first for one fixed point and then for two fixed points, emphasizing the role of the homogeneous
sector and the fixed--state conditions.  In Sec. \ref{sec: capacity eternal} we compute the capacity for one interval and for
two intervals, and we analyze the Lorentzian continuation and the resulting time dependence in the
two--QES saddle.  We conclude in Sec. \ref{conclu} with lessons and open directions, including extensions to
non-factorized regimes (full twist four--point function), evaporating backgrounds, and multi--island
configurations where the replica dynamics is expected to be even richer.


\section{Review of RST gravity}
\label{Sec: reviewRST}

The goal of this section is simply to fix conventions and collect the pieces of RST gravity that
we will use throughout the replica analysis. The RST model is a semiclassically solvable extension
of CGHS dilaton gravity in which the Polyakov anomaly can be included consistently and the
backreacted geometry can still be treated analytically~\cite{CGHS1992,RST1992,Polyakov1981,Russo:1992ht,Fabbri:2005mw}.
We work in conformal gauge and use the standard RST field redefinitions that linearize the
semiclassical equations, with normalization \(8G_N=1\) and a large-\(N\) conformal sector. Our focus
in what follows is the eternal black hole family and the corresponding generalized entropy for the
one and two-interval configurations that enter the island/QES discussion~\cite{HartmanShaghoulianStrominger2020}.
No new results are contained in this review; the point is to make the later replica equations and
their boundary conditions unambiguous, and to highlight the ingredients that matter for the capacity
in particular, namely the genuinely dynamical conformal factor and the role of global constraints.

The CGHS model is a two dimensional dilaton $\phi$ theory coupled to $N$ free scalar fields $f_i$
\begin{equation}
    S_{\text{CGHS+CFT}} = \frac{1}{2\pi}\int d^2x\sqrt{-g}e^{-2\phi}\left[R+4(\nabla\phi)^2+4\lambda^2\right] -\frac{1}{4\pi}\sum_{i=1}^N\int d^2x \sqrt{-g}(\nabla f_i)^2.
\end{equation}
This theory can be obtained by dimensional reduction of some higher dimensional string modified gravity action and $\lambda$ is the mass or charge of extremal black hole solutions \cite{Fabbri:2005mw}. CGHS vacuum solutions have black holes with mass $M$ which represents deviation from extremality of the higher dimensional solutions. In this two dimensional theory the conformal anomaly of matter fields (Hawking radiation) can be computed and the backreaction of the geometry due to this quantum contribution in the stress tensor can be tracked. The effective action from which the semiclassical equations of motion can be derived is the effective RST gravity model
\begin{align}
\begin{split}
    S_{\text{RST}}  &= S_\text{CGHS+CFT} + N(S_\text{rst}+S_\text{Polyakov}),\\
    S_\text{rst}+S_\text{Polyakov} &=-\frac{\hbar}{48\pi}\int d^2x\sqrt{-g}\left(\phi R+\frac{1}{2}R\frac{1}{\square}R\right).
\end{split}
\end{align}
The Polyakov term induces the conformal anomaly and the $S_{rst}$ term restores the solvability of the model\footnote{It was realized that $S_\text{rst}$ was needed to preserve, at the semiclassical level, the conserved current present in the classical theory $\partial^\mu(\rho-\phi)$. This term ensures the conserved current in the effective action.}. Naively the metric has $3$ degrees of freedom, adding the dilaton we have $4$ equations of motion to solve. However, due to its vanishing Weyl tensor, gravity has no dynamical degrees of freedom in two dimensions and therefore, we can gauge fix metric components and transform equations of motion into constraint equations. It is natural to invoke the conformal gauge 
\begin{equation}
    ds^2 = -e^{2\rho(x^+,x^-)}dx^+dx^-,
\end{equation}
with $g_{++}=g_{--}=0$ and $g_{+-}= -\frac{1}{2}e^{2\rho}$. The scalar curvature is $R = 8e^{-2\rho}\partial_+\partial_- \rho$, also $\sqrt{-g}=\frac{1}{2}e^{2\rho}$ and the non local term later becomes $\frac{1}{\square}R=-2\rho$. In this gauge the action is
\begin{equation}
    S_{\text{RST}} = \frac{1}{2\pi}\int dx^+dx^-\left[e^{-2\phi}\left(4\partial_+\partial_-\rho -8\partial_+\phi\partial_-\phi+2\lambda^2e^{2\rho}\right)+\frac{N\hbar}{6}(\rho-\phi)\partial_+\partial_-\rho+\sum_{i=1}^{N}\partial_+f_i \partial_-f_i\right].
\end{equation}
Matter decouples from the geometry and it only affect the gravity equations through the stress energy tensor. It turns out this theory can be solved analytically in terms of new fields variables
\begin{equation}
    \Omega = \frac{12}{N}e^{-2\phi} +\frac{\phi}{2} - \frac{1}{4}\ln\left(\frac{48}{N}\right),\qquad
    \chi=\frac{12}{N}e^{-2\phi} - \frac{\phi}{2} + \rho + \frac{1}{4}\ln\left(\frac{3}{N}\right),\label{new_dilaton}
\end{equation}
with action ($\hbar=1$)
\begin{equation}
    S = \frac{N/6}{2\pi}\int d^2x \left(\partial_+ \Omega \partial_- \Omega -\partial_+\chi \partial_- \chi + \lambda^2e^{2\chi - 2\Omega}\right) + \frac{1}{2\pi}\sum_{i=1}^N \int d^2x\partial_+f_i\partial_-f_i,
\end{equation}
and e.o.m. and constraint equations
\begin{align}\label{rst-eoms-arbitrarygauge}
\begin{split}
    \partial_+\partial_-\Omega= -\lambda^2e^{2\chi-2\Omega},\qquad 
    &\partial_+\partial_-\chi= -\lambda^2e^{2\chi-2\Omega},\\
    \partial_\pm\Omega\partial_\pm\Omega-\partial_\pm\chi\partial_\pm\chi+\partial_\pm^2\chi &=-\frac{6}{N}\sum_{i=1}^N\partial_\pm f_i\partial_\pm f_i-t_\pm. 
\end{split}
\end{align}
$t_\pm(x^\pm)$ reflects the non-local nature of the anomaly and comes from integration of conservation of the stress tensor, it is determined by boundary conditions. The first two equations make evident that $\Omega$ and $\chi$ differ only by quiral functions\footnote{This is the \textit{semiclassical} conserved current $\partial_\mu\partial^\mu(\chi-\Omega)=0$.} $\Omega = \chi +w^+(x^+)+w^-(x^-)$. It is also clear that conformal gauge is preserved under transformations $x^\pm\to z^\pm(x^\pm)$. This residual gauge invariance can be fixed to take 
\begin{equation}
        \Omega = \chi\,, \qquad \left(\phi = \rho_K - \frac{1}{2}\ln \frac{N}{12}\right),
\end{equation}
and this condition defines the Kruskal gauge, where $\rho_K$ is the conformal factor in this coordinates, and it can be obtained by solving \eqref{new_dilaton} implicitly
\begin{align}\label{conf_factor-kruskal_gauge}
\begin{split}
    \rho_K &=2\Omega +\ln 2 + \frac{1}{2}W_{-1}\left(-e^{-4\Omega}\right),
\end{split}
\end{align}
with $W_{-1}$ the product logarithm function\footnote{The -1 branch is taken so that any metric is asymptotically flat at null infinity.}. As remarked before, $\rho$ does not transform homogeneously
\begin{equation}\label{ruletransform_rho}
    \rho(z^+,z^-) \to \rho_K(x^+,x^-) = \rho\left(z^+(x^+),z^-(x^-)\right) + \frac{1}{2}\ln\left( z'^+(x^+)z'^-(x^-)\right). 
\end{equation}
Finally the gravity equations to solve in this theory are
\begin{align}\label{RST_eoms}
    \partial_+\partial_-\Omega = -\lambda^2,\qquad
    \partial_\pm^2\Omega = -T_{\pm\pm}^f-t_\pm,
\end{align}
with stress tensor
\begin{equation}
    T_{\pm\pm}^f=\frac{6}{N}\sum_{i=1}^N\partial_\pm f\partial_\pm f = \frac{12\pi}{N}\Big(T_{\pm\pm}^f\Big)_{\text{conv.}},
\end{equation}
where the relation between our $f$ fields and the conventional ones is $f=\sqrt{2\pi}f_{\text{conv}}$. We require that $\Omega \sim -x^+x^-$ at $\mathcal{I}^{\pm}$ so the solutions are asymptotically flat. As a function of $\phi$, $\Omega$ has a minimum at $\phi_{\text{min}} = -\frac{1}{2}\ln \frac{N}{48}$ and it satisfies that $\Omega \geq \frac{1}{4}$ in any physically relevant region. This ''new" dilaton should be regarded as the \textit{quantum} corrected area of the higher dimensional spheres from which this theory can emerge, and $\Omega_{\text{min}}$ is now the boundary of this space. Black hole formation and evaporation can be modeled by an incoming shockwave of energy, and in that case the dynamics of $\Omega_{\text{min}}$ usually indicates the phases of the evaporation of the black hole \cite{Fabbri:2005mw}.

Finally, it is of interest to calculate thermodynamic entropies in this theory. Typically $\frac{1}{G_{\text{eff}}}\sim \Phi$ with $\Phi$ the combined dilaton quantities multiplying the scalar curvature. The Polyakov term corrects the thermodynamic entropy of black holes and it should be regarded as the matter contribution to it \cite{BHEntropy-Myers, Fiola1994}
\begin{equation}\label{thermo-entropy-RST}
    S_{BH} = \frac{N}{6}\left(\chi(\text{Horizon})-\chi_{\text{crit}}\right), \qquad \chi_{\text{crit}}=\frac{1}{4}. 
\end{equation}

\subsection{Eternal black hole solution}
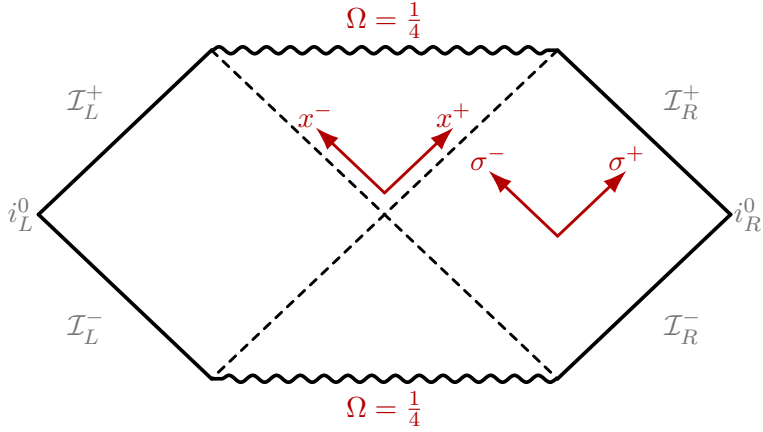
\begin{figure}[t]
\centering
\begin{tikzpicture}[
    scale=1.15,
    line cap=round,
    line join=round,
    boundary/.style={black, line width=1.4pt},
    horizon/.style={black, line width=1.1pt, dashed},
    sing/.style={black, line width=1.4pt, decorate,
                 decoration={snake, amplitude=1.3pt, segment length=9pt}},
    redlab/.style={red!70!black},
    graylab/.style={black!55},
    arr/.style={-{Latex[length=3.2mm,width=2.2mm]}, line width=1.1pt, red!70!black}
]

\coordinate (L)  at (-4, 0);
\coordinate (R)  at ( 4, 0);
\coordinate (TL) at (-2.0, 1.9) ;
\coordinate (TR) at ( 2.0, 1.9) ;
\coordinate (BR) at ( 2.0,-1.9) ;
\coordinate (BL) at (-2.0,-1.9) ;

\draw[boundary] (L) -- (TL);
\draw[boundary] (TR) -- (R);
\draw[boundary] (R) -- (BR);
\draw[boundary] (BL) -- (L);

\draw[sing] (TL) -- (TR);
\draw[sing] (BR) -- (BL);

\draw[horizon] (TL) -- (BR);
\draw[horizon] (TR) -- (BL);

\node[graylab] at ($(L)+(-0.2,0.0)$) {$i^0_{L}$};
\node[graylab] at ($(R)+( 0.2,0.0)$) {$i^0_{R}$};

\node[graylab] at ($(-3.45, 1.3)$) {${\mathcal I}^{+}_{L}$};
\node[graylab] at ($(-3.45,-1.3)$) {${\mathcal I}^{-}_{L}$};

\node[graylab] at ($( 3.45, 1.3)$) {${\mathcal I}^{+}_{R}$};
\node[graylab] at ($( 3.45,-1.3)$) {${\mathcal I}^{-}_{R}$};

\node[redlab] at ($(0, 2.25)$) {$\Omega=\tfrac{1}{4}$};
\node[redlab] at ($(0,-2.25)$) {$\Omega=\tfrac{1}{4}$};

\coordinate (C1) at (0,+0.25);
\draw[arr](C1) -- ++(-0.8,0.76) ;
\draw[arr](C1) -- ++(0.8,0.76)   ;

\node[redlab] at ($(C1)+(-0.8, 0.9)$) {$x^{-}$};
\node[redlab] at ($(C1)+(0.8, 0.9)$) {$x^{+}$};

\coordinate (C1) at (2,-0.25);
\draw[arr](C1) -- ++(-0.8,0.76) ;
\draw[arr](C1) -- ++(0.8,0.76)   ;

\node[redlab] at ($(C1)+(-0.8, 0.9)$) {$\sigma^{-}$};
\node[redlab] at ($(C1)+(0.8, 0.9)$) {$\sigma^{+}$};

\end{tikzpicture}
\caption{ Penrose diagram of the eternal RST black hole. The wavy boundaries $\Omega=1/4$ represent the singularity. The dashed lines indicate the horizons. Kruskal type coordinates and right wedge Minkowski coordinates are also shown.}
   \label{etermal-BH}
\end{figure}

The vacuum $T_{\pm\pm}^f= t_\pm=0$ black hole solution to \eqref{RST_eoms} in Kruskal coordinates is
\begin{equation}\label{eternal_BH_sol}
    \Omega(x^+,x^-) = \frac{M}{\lambda}-\lambda^2 x^+x^-,
\end{equation}
with metric given by the conformal factor \eqref{conf_factor-kruskal_gauge}. $M>\frac{1}{4}$ is the mass of the black hole and is chosen as a boundary condition in the Newtonian limit. The geometry has a spatial singularity at $\Omega=\frac{1}{4}$ and an event horizon at $x^+x^-=0$. In Fig. \ref{etermal-BH} we show the Penrose diagram for the eternal BH solution.

The geometry is asymptotically flat in the $-x^+x^-\to\infty$ limit where we have
\begin{equation}
    e^{2\rho(x^+,x^-)}\simeq -\frac{1}{\lambda^2x^+x^-}, 
\end{equation}
which is clear in right Minkowski coordinates $\lambda x^\pm = \pm e^{\pm \lambda \sigma^\pm}$ with $\sigma^\pm =t\pm y$
\begin{equation}
    ds^2 \approx \frac{dx^+dx^-}{\lambda^2x^+x^-}=-d\sigma^+d\sigma^-=-dt^2+dy^2.
\end{equation}
From the definition of the Hartle-Hawking-Israel state we can transform into Minkowski coordinates (which are inertial at infinity) and make the flux of Hawking radiation explicit 
\begin{equation}
    t_+(\sigma^+) = \langle x^+|:T_{++}(\sigma^+):|x^+\rangle = \frac{1}{4}.
\end{equation}

The temperature associated to this geometry is
\begin{equation}
    T = \frac{\lambda}{2\pi},
\end{equation}
which is independent of $M$, this implies that the canonical ensemble is ill defined\footnote{In the literature it is argued that suitable boundary counterterms in the action can take care of the issue \cite{HartmanShaghoulianStrominger2020}.} (since temperature is not a free parameter) and that heat capacity is infinite. The entropy of the black hole is given by
\begin{equation}
    S_{BH} = \frac{NM}{6\lambda}.
\end{equation}

\subsection{Generalized entropy for 1 and 2 intervals}
It is known that the island formula typically gives a unitary Page curve for the entropy of the Hawking radiation \cite{AHMST2019,HartmanShaghoulianStrominger2020, GautasonEtAl2020, Anegawa2020}. Here we briefly summarize the result for 1 and 2 intervals in RST gravity.
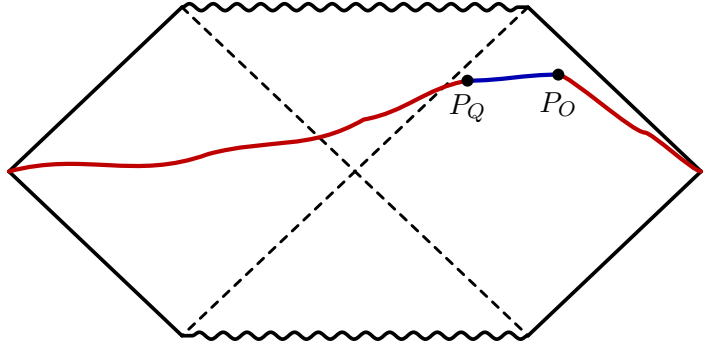
\begin{figure}[t]
    \centering
    \begin{tikzpicture}[
    scale=1.15,
    line cap=round,
    line join=round,
    boundary/.style={black, line width=1.4pt},
    horizon/.style={black, line width=1.1pt, dashed},
    sing/.style={black, line width=1.4pt, decorate,
                 decoration={snake, amplitude=1.3pt, segment length=9pt}},
    redline/.style={red!75!black, line width=1.6pt},
    blueline/.style={blue!70!black, line width=1.6pt},
    dot/.style={circle, fill=black, inner sep=1.6pt}
]

\coordinate (L)  at (-4, 0);
\coordinate (R)  at ( 4, 0);
\coordinate (TL) at (-2.0, 1.9) ;
\coordinate (TR) at ( 2.0, 1.9) ;
\coordinate (BR) at ( 2.0,-1.9) ;
\coordinate (BL) at (-2.0,-1.9) ;

\draw[boundary] (L) -- (TL);
\draw[boundary] (TR) -- (R);
\draw[boundary] (R) -- (BR);
\draw[boundary] (BL) -- (L);

\draw[sing] (TL) -- (TR);
\draw[sing] (BR) -- (BL);

\draw[horizon] (TL) -- (BR);
\draw[horizon] (TR) -- (BL);

\coordinate (A)  at (-4,0);
\coordinate (PQ) at ( 1.30, 1.05);
\coordinate (PO) at ( 2.35, 1.12);
\coordinate (B1) at ( 3.35, 0.45);
\coordinate (B)  at ( 4, 0);

\draw[redline]
  (A)
    to[out=15,in=200] (-1.70,0.20)
    to[out=15,in=210] ( 0.10,0.6)
    to[out=10,in=190] (PQ);

\draw[blueline]
  (PQ) to[out=0,in=180] (PO);

\draw[redline]
  (PO)
    to[out=-5,in=180,looseness=0.3] (B1)
    to[out=-5,in=180,looseness=0.3] (B);

\node[dot] at (PQ) {};
\node[dot] at (PO) {};
\node[above] at ($(PQ)+(0,-0.6)$) {$P_Q$};
\node[above] at ($(PO)-(0,0.6)$) {$P_O$};

\end{tikzpicture}
   \caption{Single-interval configuration in the eternal RST black hole. The radiation region is anchored at $P_O$ on the asymptotic boundary (red curve). In the island saddle the generalized entropy (and capacity) is evaluated using a single quantum extremal point $P_Q$, with the exterior and island segment shown in red and the complement segment in blue.}
    \label{conf_isla}
\end{figure}

First we start with the interval $\text{Rad}=[P_O,P_\Lambda]$ taking $P_\Lambda\to i^0_R$.  The prescription is to minimize the generalized entropy, allowing the contribution of an island region $\mathcal{I}$
\begin{equation}
S_{\text{gen}}(\mathcal{I}\cup\text{Rad})
=\operatorname*{ext}_{\partial\mathcal{I}}
\left\{\frac{\text{Area}(\partial \mathcal{I})}{4G_{\text{eff}}}
+S_{vN}(\mathcal{I}\cup \text{Rad})\right\},
\end{equation}
with $S_{vN}$ the entanglement entropy (EE) of $N$ massless scalar fields in a conformally flat geometry. Since we are taking $P_\Lambda\to i^0_R$, and assuming the global state to be pure we can consider the complement region $\overline{\mathcal{I}\cup\text{Rad}}=[P_Q,P_O]=B$
\begin{equation}
    S_{\text{gen}}(B) = \frac{\text{Area}(P_Q)}{4G_{\text{eff}}} +S_{vN}(B), 
\end{equation}
as shown in Fig. \ref{conf_isla}. At this point we need to understand what is $\text{Area}(P_Q)$ in this situation. As remarked before, the formula \eqref{thermo-entropy-RST} has a matter contribution, so when applying the formula in this context we must subtract the Polyakov contribution. This term is already taken into account in the conformal factor coming from the EE. Alternatively we can use \eqref{thermo-entropy-RST} without including the conformal factor at $P_Q$ in $S_{vN}$. Then we extremize
\begin{equation}
    S_{\text{gen}}(B) = \frac{N}{6}\left(\Omega(P_Q)-\Omega_{\text{crit}}\right)+\frac{N}{6}\ln \frac{-(x^+_Q-x^+_O)(x^-_Q-x^-_O)}{\epsilon e^{-\rho_K(P_O)}},
\end{equation}
with respect to $x_Q^\pm$ for the eternal black hole background. The matter EE comes with the usual UV cutoff $\epsilon$ that regularizes its divergence. The equations 
\begin{equation}
\partial_\pm S_{\text{gen}}(B)=x^\mp_Q+\frac{1}{x_Q^\pm-x_O^\pm}=0  ,
\end{equation}
always has a quantum extremal surface solution $P_Q$. This is expected since the island region purifies the infrared divergent EE induced by $P_\Lambda$ in $\text{Rad}$. We consider the limit $x^+_O\to +\infty$, in which case the position of the QES is
\begin{equation}
    x_Q^\pm \approx -\frac{1}{x^\mp_O}.
\end{equation}
This means that $P_Q$ sits outside the event horizon and lies on it when $P_O\in \mathcal{I}^+$. The generalized entropy gives
\begin{equation}\label{s_gen}
    S_{\text{gen}}(B) = \frac{NM}{6\lambda} + \frac{N}{12}\ln \frac{-x_O^+x_O^-}{\epsilon^2} = \frac{NM}{6\lambda} + \frac{N}{6}\lambda y_O-\frac{N}{6}\ln \lambda\epsilon,
\end{equation}
where we dropped all terms of order $\mathcal{O}(1)$ in $M,x^+_O$ and $1/\epsilon$.

\begin{figure}[t]
\centering
\begin{tikzpicture}[
    line cap=round,
    line join=round,
    boundary/.style={black, line width=1.2pt},
    horizon/.style={black, line width=1.0pt, dashed},
    sing/.style={black, line width=1.2pt, decorate,
                 decoration={snake, amplitude=1.2pt, segment length=8pt}},
    redline/.style={red!75!black, line width=1.4pt},
    blueline/.style={blue!70!black, line width=1.4pt},
    dot/.style={circle, fill=black, inner sep=1.5pt},
    lab/.style={font=\normalsize}
]

\begin{scope}[scale=0.80]
  \coordinate (L1)  at (-4, 0);
\coordinate (R1)  at ( 4, 0);
\coordinate (TL1) at (-2.0, 1.9) ;
\coordinate (TR1) at ( 2.0, 1.9) ;
\coordinate (BR1) at ( 2.0,-1.9) ;
\coordinate (BL1) at (-2.0,-1.9) ;

  \draw[boundary] (L1) -- (TL1);
  \draw[boundary] (TR1) -- (R1);
  \draw[boundary] (R1) -- (BR1);
  \draw[boundary] (BL1) -- (L1);

  \draw[sing] (TL1) -- (TR1);
  \draw[sing] (BR1) -- (BL1);

  \draw[horizon] (TL1) -- (BR1);
  \draw[horizon] (TR1) -- (BL1);

  \coordinate (POl1) at (-3,0.8);
  \coordinate (POr1) at (3,0.8);

  \draw[redline] (L1) -- (POl1);
  \draw[redline] (POr1) -- (R1);

  \node[dot] at (POl1) {};
  \node[dot] at (POr1) {};
  \node[lab, above left]  at (POl1) {$P_{\bar{O}}$};
  \node[lab, above right] at (POr1) {$P_O$};
\end{scope}

\begin{scope}[xshift=7.9cm, scale=0.80]
\coordinate (L2)  at (-4, 0);
\coordinate (R2)  at ( 4, 0);
\coordinate (TL2) at (-2.0, 1.9) ;
\coordinate (TR2) at ( 2.0, 1.9) ;
\coordinate (BR2) at ( 2.0,-1.9) ;
\coordinate (BL2) at (-2.0,-1.9) ;

  \draw[boundary] (L2) -- (TL2);
  \draw[boundary] (TR2) -- (R2);
  \draw[boundary] (R2) -- (BR2);
  \draw[boundary] (BL2) -- (L2);

  \draw[sing] (TL2) -- (TR2);
  \draw[sing] (BR2) -- (BL2);

  \draw[horizon] (TL2) -- (BR2);
  \draw[horizon] (TR2) -- (BL2);

  \coordinate (POl2) at (-2.6,1.1);
  \coordinate (POr2) at (2.6,1.1);

  \coordinate (PQl2) at (-1.4, 1.1);
  \coordinate (PQr2) at ( 1.4, 1.1);

  \draw[redline] (L2) -- (POl2);
  \draw[redline] (POr2) -- (R2);

  \draw[blueline] (POl2) to[out=35,in=200] (PQl2);
  \draw[blueline] (PQr2) to[out=-20,in=160] (POr2);

  \draw[redline] (PQl2) -- (PQr2);

  \node[dot] at (POl2) {};
  \node[dot] at (POr2) {};
  \node[dot] at (PQl2) {};
  \node[dot] at (PQr2) {};

  \node[lab, above left]  at (POl2) {$P_{\bar O}$};
  \node[lab, above right] at (POr2) {$P_O$};

  \node[lab, above] at  (-1.4, 0.2) {$P_{\bar Q}$};
  \node[lab, above] at ( 1.4, 0.2) {$P_Q$};
\end{scope}

\end{tikzpicture}
\caption{Two-interval configuration in the eternal RST black hole. Left: the radiation region is anchored at two boundary points $P_{\bar O}, P_O$ (one on each asymptotic boundary), no-island configuration. Right: in the island saddle the entanglement wedge includes a symmetric island bounded by two quantum extremal points $[P_{\bar Q},P_Q]$; the generalized entropy is evaluated on the red segments (exterior + island), which gives the same as for the blue segments (assuming global pure state and Haag duality).}
 \label{fig:2QES}
\end{figure}
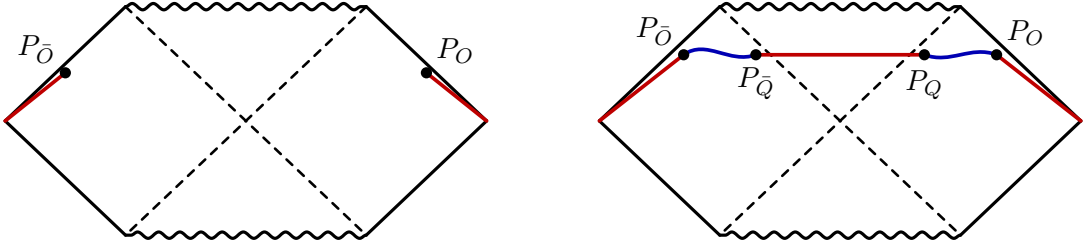

For two intervals the initial configuration is $\overline{\text{Rad}}\cup \text{Rad}=[P_{\bar{\Lambda}},P_{\bar{O}}]\cup [P_O,P_\Lambda]$. All barred points are the mirrored versions on the unbarred, which in euclidean coordinates correspond to $\tau \to -(\tau +\frac{\pi}{2})$. Again, we take the limit $P_\Lambda \to \mathcal{I}^+_R$ as shown in Fig. \ref{fig:2QES}. Now we take the generalized entropy to include the area term for two-QES $P_Q$ and $P_{\bar{Q}}$\footnote{By symmetry arguments it can be shown that the left QES is the mirrored point to $P_Q$.}, and the EE is again, by Haag duality, the EE of $[P_{\bar{O}},P_{\bar{Q}}]\cup [P_Q,P_O] = \bar{B}\cup B$. As is known\footnote{To mimic island dynamics in the minimization of the generalized entropy in the eternal black hole background it is necessary to take $S_{vN}$ normalized at some initial time $t=0$ for which $S_{vN}=0$.}, the minimization of the generalized entropy is not always realized by the inclusion of the island region $\mathcal{I}=[P_{\bar{Q}},P_Q]$ (either because there is no solution to $P_Q$ and $P_{\bar{Q}}$ as functions of $P_O$ and $P_{\bar{O}}$ or simply because the no--island configuration has lower generalized entropy). At early times the generalized entropy gives 
\begin{equation}\label{equation:s_gen_2int_earlyTimes}
    S_{\text{gen}}(\overline{\text{Rad}}\cup\text{Rad}) = S_{vN}[P_{\bar{O}},P_O] \approx \frac{N}{3}\lambda t_O- \frac{N}{3}\ln \lambda\epsilon,
\end{equation}
which grows linearly with time.
When considering the factorization regime (late times) in which $S_{vN}(\bar{B}\cup B) = S_{vN}(\bar{B})+S_{vN}(B)$, the island inclusion minimizes the generalized entropy an its boundaries approaches the past and future event horizon. For the entropy we get
\begin{equation}
   S_{\text{gen}}(\overline{\text{Rad}}\cup \mathcal{I}\cup\text{Rad})= S_{\text{gen}}(\bar{B}\cup B) = 2\left(\frac{NM}{6\lambda}+ \frac{N}{6}\lambda y_O-\frac{N}{6}\ln \lambda \epsilon\right)  = 2S_{\text{gen}}(B).
\end{equation}
This result is time independent (since the black hole is not evaporating in this setup) and represents the unitarity aspect of the Page curve.


\section{Replica equations and Capacity of entanglement}
\label{sec:replica-eqs-CoE}
In this section we set up the gravitational replica construction for our interval configurations
and solve the semiclassical RST equations in the neighborhood of \(n=1\). The strategy is to treat
the \(\mathbb{Z}_n\) orbifold as a small deformation of the physical geometry, expand the fields
and the on-shell action in powers of \((n-1)\), and determine the resulting backreaction by
imposing both single valuedness around the branch points and the correct asymptotic state at the
boundaries. The key point is that the linearized equations contain homogeneous integration constants that are
invisible in a purely local analysis but are uniquely fixed once the appropriate global conditions
(equivalently, the asymptotic charges defining the state) are imposed; in our problem, the capacity
is precisely sensitive to this globally fixed data.

The replica trick computes the entanglement entropy $S(A)$ by means of the Rényi Entropy $S_n(A)$ and a limit on the analytic continuation of the replica parameter $n\in \mathbb{N}$
\begin{equation}
    S(A) = \lim_{n\to1}S_n(A) = \lim_{n\to 1}\frac{1}{1-n} \log \left(\text{Tr}[\rho_A^n]\right).
\end{equation}
The path integral representation of a density matrix relates the Rényi Entropy to a partition function on a replicated manifold
\begin{equation}
    \text{Tr}[\rho_A^n] = Z[\widetilde{\mathcal{M}}_{n}],
\end{equation}
where $\widetilde{\mathcal{M}}_n$ is $n$ copies of the original space $\mathcal{M}$ glued together in a cyclical way. It encodes in its geometry and topology  details of $\mathcal{M}$ and $A$. From this it follows that
\begin{equation}
    S(A) = -\partial_n\left. \left(\frac{\log Z_n}{n}\right)\right\vert_{n=1}.
\end{equation}
In special cases this formalism recovers well known results of EE when $\mathcal{M}$ is flat or conformally flat, with fixed metric. In this context the capacity of entanglement $C(A)$ can be defined as the $n\to1$ limit of a derivative of a modified Rényi Entropy $\tilde S_n$
\begin{equation}
    C^{(n)} \equiv -n \partial_n \widetilde{S} ^{(n)}\hspace{.5cm}  \text{with}\hspace{.5cm} \widetilde{S}^{(n)} =\partial_{\frac{1}{n}} \left(\frac{1}{n}\log \text{Tr}\rho_{A}^n\right).
\end{equation}
Alternatively it can be expressed as
\begin{equation}
    C(A) = \lim_{n\to1} n^2 \partial^2_n \log \text{Tr}\rho^n_{A}. 
\end{equation}
This shows that CoE generally needs more information about the replica procedure due to the second derivative in $n$.  These definitions are motivated by their thermodynamical counterparts as shown in table \ref{tab:analogy}.
\begin{table}[t]
\centering
\begin{tabular}{r l c r l}
\toprule
\multicolumn{2}{c}{\textbf{Statistical mechanics}} & \quad & \multicolumn{2}{c}{\textbf{R\'{e}nyi analogue}} \\
\midrule
Inverse temperature: & $\beta$ & & Replica parameter: & $n$ \\[4pt]
Hamiltonian: & $H$ & & Modular Hamiltonian: & $H_A = -\log \rho_A$ \\[4pt]
Partition function: & $Z(\beta) = \mathrm{Tr}\left[e^{-\beta H}\right]$ & & Replica partition function: & $Z(n) = \mathrm{Tr}_A\left[e^{-n H_A}\right]$ \\[4pt]
Free energy: & $F(\beta) = -\beta^{-1} \log Z(\beta)$ & & Replica free energy: & $F(n) = -n^{-1} \log Z(n)$ \\[4pt]
Energy: & $E(\beta) = -\partial_\beta \log Z(\beta)$ & & Replica energy: & $E(n) = -\partial_n \log Z(n)$ \\[4pt]
Thermal entropy: & $S(\beta) = \beta^2\, \partial_\beta F(\beta)$ & & Refined R\'{e}nyi entropy: & $\tilde{S}^{(n)} = n^2\, \partial_n F(n)$ \\[4pt]
Heat capacity: & $C(\beta) = -\beta\, \partial_\beta S(\beta)$ & & Capacity of entanglement: & $C^{(n)} = -n\, \partial_n \tilde{S}^{(n)}$ \\
\bottomrule
\end{tabular}
\caption{An analogy between statistical mechanics and R\'{e}nyi entropic quantities \cite{KawabataEtAl2105}. Here $n$ denotes
the replica parameter.}
\label{tab:analogy}
\end{table}

When $\mathcal{M}$ is curved, the partition function on $\widetilde{\mathcal{M}}_n$ is a gravitational path integral which will be treated by the saddle point approximation. Each saddle will inherit a replica symmetry with fixed points $\Sigma$ under $\mathbb{Z}_n$, the permutation of copies. From this it follows that
\begin{align}
\begin{split}
    \log Z[\widetilde{\mathcal{M}}_n] &=  -I_{\text{grav}}[\widetilde{\mathcal{M}}_n] + \log Z_{\text{matter}}[\widetilde{\mathcal{M}}_n]\\
    &= -n\left(I_{\text{grav}}[\mathcal{M}_n] + T_n\int_{\Sigma}d^{d-2}y\sqrt{|\gamma|}\right)+ \log Z_{\text{matter}}[\widetilde{\mathcal{M}}_n],
\end{split}
\end{align}
with $\mathcal{M}_n = \widetilde{\mathcal{M}}_n/\mathbb{Z}_n$ and $T_n=\frac{1}{4G_{\text{eff}}}(1-\frac{1}{n})$ a tension like parameter given to the surface $\Sigma$, parametrized by coordinates $y$ and with induced metric $\gamma_{\mu\nu}$. Coupling gravity to this brane induces the conical defects in $\mathcal{M}_n$ needed for $\widetilde{\mathcal{M}}_n$ to be smooth. We define the area of $\Sigma$ by
\begin{equation}
    \mathcal{A}^{(n)} = \frac{1}{4G_{\text{eff}}}\int_\Sigma d^{d-2}y\sqrt{|\gamma|}.
\end{equation}
The modified Rényi entropy takes the form
\begin{equation}
    \widetilde{S}^{(n)} = \mathcal{A}^{(n)} +\widetilde{S}^{(n)}_{\text{matter}},
\end{equation}
and therefore
\begin{equation}
    C^{(n)} =C^{(n)}_{\text{matter}} -n\partial_n \mathcal{A}^{(n)}.
\end{equation}
This is a simple expression and in general very difficult to calculate explicitly, mainly because we need to solve the semiclassical gravity equations of motion in $\mathcal{M}_n$ in the presence of conical singularities and twist operator insertions at order $n-1$. One important remark is that we need first to extremize $S_{\text{gen}}$ to choose in what topology (how many QES) we should solve the equations.

In RST the area term is given by the dilaton-dependent combination that appears in \(\chi\), and, to linear order around \(n=1\), the
replica variation of the gravitational contribution reduces to evaluating \(\delta\chi\) at the
\(\mathbb{Z}_n\) fixed points. As a result, the generalized capacity can be written in a compact
form as the matter contribution minus the area variation at each quantum extremal surface. For a configuration with $N_{QES}$ components is simply
\begin{equation}
C_{\rm gen}
= C_{\rm matter}
-\frac{N}{6}\sum_{a=1}^{N_{\rm QES}}
\delta\chi\Big|_{Q_a}\,,
\label{eq:Cgen_multiQES}
\end{equation}
where the evaluation is on the regularized position of the QES $Q_a$. The remainder of this section is devoted to determining \(\delta\chi\)
globally: while its singular structure near each QES is fixed locally, the homogeneous
modes are fixed only after imposing the appropriate boundary conditions at infinity. We note that in using the semiclassical area we must consider, as we did for the entropy, the matter contribution without conformal factors on quantum extremal surfaces.

\subsection{RST replica equations of motion - 1 interval}\label{rst1int}
The gravitational replica trick computes \(\Tr \rho_A^{\,n}\) in a fixed boundary state: we are not defining a variational problem with freely adjustable asymptotics, but rather evaluating the replicated path integral with the same asymptotic data that define \(\rho_A\). In the \(\mathbb{Z}_n\)-quotient orbifold description, this implies that the quotient geometry must preserve the boundary charges that characterize the state. In particular, any parameter that labels nonequivalent semiclassical states cannot be allowed to drift when passing to the orbifold. For the eternal RST black hole this means that the ADM mass must be kept fixed,
\begin{equation}
\tilde M = M\,,
\label{eq:state_fixing_M}
\end{equation}
and we will use this condition to fix the otherwise ambiguous homogeneous (constant) mode that appears in the global solution at order \((n-1)\).

Going to the orbifold, on the other hand, alter the local thermal identification near the branch point. Taking the quotient reduces the length of the thermal circle by a factor of $n$, which is implemented in the orbifold configurations by rescaling $\tilde{\beta} = 2\pi/\lambda$ to $\beta = 2\pi/n\lambda$. In a black hole saddle this is naturally encoded as a rescaling of the surface gravity, \(\tilde\kappa = n\,\kappa\). In RST (as in CGHS), the parameter \(\lambda\) sets the relevant thermal scale of the eternal solution, and it is convenient to implement the orbifold thermality by
\begin{equation}
\lambda \to n\,\lambda\,.
\label{eq:lambda_tilde_equals_nlambda}
\end{equation}
Operationally, \eqref{eq:state_fixing_M} removes the residual freedom associated with the constant mode of the global solution (for instance, the combination that would shift the mass in our linearized parametrization), leaving a unique global solution consistent with the replicated problem.

While Sec. \ref{Sec: reviewRST} was phrased in Lorentzian terms, the replica construction that defines \(\Tr\rho_A^n\) is naturally formulated as a Euclidean path integral, and therefore we perform the Wick rotation  \(t\to -i\tau\) and use complex conformal coordinates
$z=e^{w}$ with $w=y+i\tau$ in the orbifold cone geometry with opening angle \(2\pi/n\) at each fixed point. Dependence on $z$ and $\bar{z}$ will not be made explicit to avoid clutter.  Lorentzian time dependence will be recovered at the end by analytic continuation back to real time, after the
replica derivatives have been taken.
In RST gravity with conical singularities we have 
\begin{align}
\begin{split}
    I_{\text{RST}+\delta}&= I_{\text{grav}}[\mathcal{M}_n] + T_n\int_{\Sigma}d^{d-2}y\sqrt{|\gamma|} \\
    &=-\frac{N/6}{2\pi}\int dzd\bar{z}\left\{\partial\chi\bar{\partial}\chi-\partial\Omega \bar{\partial}\Omega+ e^{2\chi-2\Omega}-\pi(1-1/n)\chi\delta(z-z_Q)\right\}.
\end{split}
\end{align}
For the CoE we only need to solve \eqref{rst-eoms-arbitrarygauge} at order $n-1$ (arbitrary $n$ solutions for one-QES was explored in App. \ref{app:U1-ansatz}), so we expand all fields and the stress energy tensor
\begin{equation}
    \Omega^{(n)}=\Omega^{(1)}+(n-1)\delta\Omega ,\qquad\chi^{(n)}=\chi^{(1)}+(n-1)\delta\chi , \qquad -\frac{6}{N}\delta T^f_{zz} = \frac{1/2}{(z-z_Q)^2} -\frac{A}{z-z_Q}, 
\end{equation}
with $A = \frac{6}{N}\left.\partial_zS^{\text{flat}}_{vN}[z,z_O]\right|_{z=z_Q}$. Thus they give
\begin{align}\label{orbifold_eqs}
\begin{split}
    \partial\bar{\partial}(\delta\Omega-\delta\chi) &= \frac{\pi}{2}\delta(z-z_Q),\\
\partial\bar{\partial}\delta\Omega&= 2(\delta\chi-\delta\Omega),\\
2\partial\Omega^{(1)}\partial(\delta\Omega-\delta\chi)+\partial^2\delta\chi&= \frac{1/2}{(z-z_Q)^2}-\frac{A}{z-z_Q}, 
\end{split}
\end{align}
The first equation implies
\begin{equation}
    \delta\Omega-\delta\chi = \frac{1}{2}\ln |z-z_Q|^2 + w_1(z)+w_2(\bar{z}),
\end{equation}
and we have a residual conformal gauge symmetry that allows to set $w_1+w_2=0$.  The second equation then reads
\begin{equation}
    \partial\bar{\partial}\delta\Omega = -\ln|z-z_Q|^2,
\end{equation}
which is solved by
\begin{equation}
    \delta\Omega = 2|z-z_Q|^2 -|z-z_Q|^2 \ln|z-z_Q|^2 + h(z)+g(\bar{z}),
\end{equation}
with $h$ and $g$ holomorphic and antiholomorphic functions. Lastly, the constraint equation can be written as

\begin{equation}
\frac{\partial\Omega^{(1)}- \overline{z-z_Q}+A}{z-z_Q}+h''=0.
\end{equation}
Using that $\Omega^{(1)} = M+z\bar{z}$  the numerator of the simple pole gives
\begin{equation}
    \bar{z}_Q +\frac{1}{z_Q-z_O},
\end{equation}
which is exactly $\partial_{z_{Q}} S_{\text{gen}}$. Imposing this number to vanish is to impose the extremization of the generalized entropy. Finally, we have that $h$ and $g$ are at most linear in $z-z_Q$
\begin{equation}
    h(z) = h_0+h_1(z-z_Q),\qquad
g(\bar{z})=g_0+g_1\overline{z-z_Q}.
\end{equation}
Then the general solution to the global equations are (restoring $\lambda^2$)
\begin{align}\label{solution1QES}
\begin{split}
    \delta\Omega &= a+b(z-z_Q)+c\overline{(z-z_Q)}+2\lambda^2|z-z_Q|^2-\lambda^2|z-z_Q|^2\ln\lambda^2|z-z_Q|^2,\\
    \delta\chi &= \delta\Omega - \frac{1}{2}\ln \lambda^2|z-z_Q|^2.
\end{split}
\end{align}
Implementing the state-fixing conditions explained at the beginning of the section yields
\begin{equation}
b=c=0,
\qquad
a=-\frac{M}{\lambda}.
\label{eq:abcd_fix_two}
\end{equation}
This completes the construction of the replica correction at order \((n-1)\): the local
singular behavior near each \(\mathbb{Z}_n\) fixed point is fixed by regularity on the cover, while
the remaining homogeneous integration constants are uniquely determined by the global conditions
that keep the asymptotic state fixed. In the next section we will use this solution to evaluate
the replicated on-shell action and extract the generalized capacity, making explicit how this
globally fixed data enters an observable that is sensitive to the correction.

\subsection{RST replica equations of motion - 2 intervals}
\label{app:rst-replica-2interval}

Here we extend the \(n\simeq 1\) orbifold analysis to the two-interval configuration relevant for
Fig.~\ref{fig:2QES}: the region \(B\cup\bar B = [P_{\bar{O}},P_{\bar{Q}}]\cup [P_Q,P_O]\)  (the complement of the island + bath regions).  On the orbifold this saddle
has two \(\mathbb{Z}_n\) fixed points,
\begin{equation}
\Sigma=\{z_1,z_2\},\qquad z_1=z_{Q},\qquad z_2=z_{\bar{Q}},
\end{equation}
in a complex coordinate \(z\).  We work to first order in \((n-1)\) throughout, and emphasize only the
new features relative to the one fixed point case.

With two fixed points, the area term will have two delta contributions, and the equations therefore will be

\begin{align}\label{2qes_replicaeqs}
\begin{split}
    \partial\bar\partial(\delta\Omega-\delta\chi)
&=\frac{\pi}{2}\Big[\delta(z-z_1)+\delta(z-z_2)\Big],
\\
\partial\bar\partial\,\delta\Omega
&=2(\delta\chi-\delta\Omega),\\
2\,\partial\Omega^{(1)}\partial(\delta\Omega-\delta\chi)+\partial^2\delta\chi
&=-\frac{6}{N}\delta T^{f}_{zz}\,.
\end{split}
\end{align}
The first equation integrates to
\begin{equation}
\delta\Omega-\delta\chi
=\frac{1}{2}\ln|z-z_1|^2+\frac{1}{2}\ln|z-z_2|^2,
\label{eq:two_diff_solution}
\end{equation}
after fixing the same residual freedom in the \(n\)-direction as in the one fixed point case.
The second equation in \eqref{2qes_replicaeqs} is then solved by a sum of particular solutions plus a single
pair of holomorphic/antiholomorphic functions,
\begin{equation}
\delta\Omega
=
\delta\Omega_1+\delta\Omega_2+h(z)+g(\bar z),
\qquad
\delta\Omega_i
=
2|z-z_i|^2-|z-z_i|^2\ln|z-z_i|^2.
\label{eq:two_omega_solution}
\end{equation}

For two intervals the CFT input is a four-point twist correlator, and the simple-pole data is
generically coupled through the cross ratio.  Here we restrict to the late-time/small-cross-ratio
(factorization) regime appropriate to the island saddle in Fig. \ref{fig:2QES}, where
\begin{equation}
S_{vN}\!\left([ P_{\bar{O}},P_{\bar{Q}}]\cup[P_Q,P_{O}]\right)
\;\approx\;
S_{vN}([P_{\bar{O}},P_{\bar{Q}}]) + S_{vN}([P_Q,P_O]).
\label{eq:factorization_assumption}
\end{equation}
As discussed in Apprendix \ref{app Tmatter}, in this regime and at order $(n-1)$ we have
\begin{equation}
-\frac{6}{N}\,\delta T^{\rm flat}_{zz}
=
\sum_{i=1}^2\left[
\frac{1/2}{(z-z_i)^2}-\frac{A_i}{z-z_i}
\right].
\label{eq:Tzz_two}
\end{equation}
In the factorization regime we have
\begin{equation}
A_1=\frac{1}{z_1-z_O},\qquad A_2=\frac{1}{z_2- z_{\bar{O}}},
\label{eq:Ai_factorized_explicit}
\end{equation}
with the labeling of endpoints in each interval defined before.

Substituting in the constraint equations it reduces to
\begin{equation}
h''(z)
+
\frac{\bar z_1+A_1}{z-z_1}
+
\frac{\bar z_2+A_2}{z-z_2}
=0.
\label{eq:hpp_two}
\end{equation}
Requiring single-valuedness for $h$ forces
\begin{equation}
\bar z_1+A_1=0,
\qquad
\bar z_2+A_2=0,
\label{eq:two_QES_conditions}
\end{equation}
which are precisely the two-QES conditions in orbifold language. The global solution can be written as (restoring $\lambda^2$)
\begin{align}\label{eq:deltaChi_global_two}
\begin{split}
    \delta\Omega
&=
a+ b z + c\bar z
+\sum_{i=1}^2\Big(2\lambda^2|z-z_i|^2-\lambda^2|z-z_i|^2\ln\lambda^2|z-z_i|^2\Big),
\\
\delta\chi
&=
\delta\Omega
-\frac12\ln\lambda^2|z-z_1|^2-\frac12\ln\lambda^2|z-z_2|^2.
\end{split}
\end{align}
As for the one-QES case, fixing global charges fixes $a,b$ and $c$ to
\begin{equation}
b=c=0,
\qquad
a=-\frac{M}{\lambda},
\qquad
\label{eq:abcd_fix_two}
\end{equation}
so the two-QES replica deformation does not introduce an independent homogeneous excitation.

This is the global solution for the two-QES in the orbifold in the regime of factorization. For a more refined calculation the full two interval twist correlation function should be used in the right hand side of the constraint equations, see App. \ref{app Tmatter} for details.


\section{Capacity of entanglement in Eternal Black Holes}\label{sec: capacity eternal}

To compute the CoE for the configuration that minimized the generalized entropy, we need to find the variation of the fields when the replica parameter $n$ deviates slightly from $n=1$. This is to say, we need to solve the replica wormholes at order $n-1$, and this statement means to solve the gravity semiclassical equations of motion in a space with conical singularities  sourced by matter with twist operator insertions. In Sec. \ref{sec:replica-eqs-CoE} we followed this procedure and found the global solutions for two different saddles (the one and two-QES configurations) in the eternal black hole background. Here we do the final computation using \eqref{eq:Cgen_multiQES}.

\subsection{Single-interval capacity (one-QES)}
We begin with the single-interval configuration of Fig. \ref{conf_isla}, where the relevant region is
\(B=[P_Q,P_O]\) and the replica saddle contains a single fixed point at \(z=z_Q\).
The general structure of the answer follows from the on-shell decomposition reviewed in Sec. \ref{sec:replica-eqs-CoE}:
the generalized capacity is the matter contribution evaluated on the orbifold, minus a purely
gravitational term coming from the localized ``area'' coupling at the fixed point. 
The matter piece is sensitive to the fact that the conformal factor is \(n\)-dependent: although this dependence drops out of the von Neumann entropy, it contributes to the capacity through derivatives on $n$ (see App. \ref{Cnmatter} for a self-contained derivation).

Using \eqref{solution1QES} we have that
\begin{align}
\begin{split}
    C_{\text{gen}}(B) &= C_{\text{matter}}(B) -\frac{N}{6}\delta\chi|_{z=z_Q+\epsilon}\\
        &= \frac{N}{6}\ln \frac{|z_O|}{\epsilon}+\frac{N}{3}\ln |z_O|-\frac{N}{6}\left.(\delta\Omega -\frac{1}{2}\ln|z-z_Q|^2)\right|_{z=z_Q+\epsilon}\\
    &= \frac{NM}{6}+\frac{N}{2}\ln |z_O| = \frac{NM}{6\lambda} +\frac{N}{2}\lambda y_O\,.
\end{split}\label{C1QES}
\end{align}
A few comments are in order. First, the one-interval generalized capacity is time independent, in parallel with the generalized
entropy on the corresponding one-QES saddle. This is consistent with the physical picture of the
eternal geometry: 
the eternal black hole is U(1) euclidean time symmetric, and the replica construction for one interval does not break the symmetry (by mapping $z_Q$ and $z_O$ to 0 and infinity respectively, see App. \ref{app:U1-ansatz} for more details). This makes $C_{\text{gen}}$ and also $S_{\text{gen}}$ time independent.

Second, we note that the generalized entropy and the capacity of entanglement has a common gravitational contribution $NM/6\lambda$, that we fixed with global charge arguments in the orbifold solution. This seems to be a common feature for both RST and JT gravity \cite{KawabataEtAl2105}, in which the area of the fixed point enters both in $S_{\text{gen}}$ and $C_{\text{gen}}$. We also note that the other gravitational contribution to the capacity makes it  UV divergence free, in contrast with the entanglement entropy.

\subsection{Two intervals capacity: early and late-time regime}\label{subsec:two-interval_capacity}
Now we consider the two-interval configuration of Fig. \ref{fig:2QES} (left), with relevant region $\overline{\text{Rad}}\cup \text{Rad}$ for which the dominant replica saddle has no fixed points at early times (the Hawking saddle) and then two fixed points after the Page time (the wormhole saddle) at euclidean positions $z_1$ and $z_2$. This is the first setup in which the capacity
becomes genuinely more informative than the generalized entropy: while the von Neumann answer on
the two-QES configuration is essentially additive in the two components, the capacity probes the detailed
\(n\)-dependence of the replica geometry and is therefore sensitive to how the two conical defects
``talk'' to each other through the global solution.

At early times $t_O<t_{\rm Page}$ the no-island configuration dominates and with no gravitational/area term in \eqref{eq:Cgen_multiQES} the generalized capacity reduces
entirely to the matter contribution,
\begin{equation}
  C_{\rm gen}^{(0)}\!\left(\overline{\rm Rad}\cup{\rm Rad}\right)
  = C_{\rm matter}\!\left(\overline{\rm Rad}\cup{\rm Rad}\right).
\end{equation}
Moreover, for the vacuum CFT in this kinematics the matter capacity coincides with the matter
von Neumann entropy \eqref{equation:s_gen_2int_earlyTimes}.

In the two-QES configuration the computation parallels the one-interval case but with one crucial new ingredient. Since the orbifold contains two conical singularities the gravitational contribution receives an interaction-type term depending on $\Delta = |z_1-z_2|$. 
In Lorentzian signature this separation becomes time dependent 
\(\Delta\to \Delta_L(t)\), and it will produce time dependence
in the generalized capacity even in regimes where \(S_{\rm gen}(t)\) has already saturated. For $\bar{B}\cup B=[P_{\bar{O}},P_{\bar{Q}}]\cup [P_{Q},P_{O}]$ we have that
\begin{align}
\begin{split}
    C_{\text{gen}}(\bar{B}\cup B) &= C_{\text{matter}}(\bar{B}\cup B) -\frac{N}{6}\left(\delta\chi|_{z=z_1+\epsilon}+\delta\chi|_{z=z_2+\epsilon}\right)\\
    &= 2C_{\text{matter}}(B) -\frac{N}{6}(\delta\chi(z_1)+\delta\chi(z_2)),
\end{split}
\end{align}
where we assume the factorization regime. Using the global solution \eqref{eq:deltaChi_global_two} we get
\begin{equation}
    \delta\chi(z_1)+\delta\chi(z_2) = \left(-2M+ 2(2|z_1-z_2|^2-|z_1-z_2|^2\ln |z_1-z_2|^2)-2(\ln \epsilon + \ln |z_1-z_2|)\right),
\end{equation}
and therefore the result is 
\begin{equation}
    C_{\text{gen}}(\bar{B}\cup B) = 2C_{\text{gen}}(B)-\frac{N}{3}\left(2\Delta^2-\Delta^2\ln \Delta^2-\ln \Delta\right)\label{82},
\end{equation}
with $\Delta = |z_1-z_2|$.  The QES are placed at $z_1 = 1/\bar{z}_O$ and $z_2 = 1/\bar{z}_{\bar{O}}$ and since $P_{\bar{O}}$ and $P_O$ are mirrored points\footnote{The right and left wedges are defined as $x^\pm=\pm e^{\pm \sigma^\pm_R }$ and $x^\pm = \mp e^{\mp \sigma_L^\pm}$ with $\sigma_R^\pm = t_R\pm y_R$ and $\sigma_L^\pm = t_L\mp y_L$ as is common in the literature.} we have the relation $z_{\bar{O}}= -\bar{z}_O$. This give 
\begin{equation}
\Delta = |z_1-z_2| = \left|\frac{1}{\bar{z}_O}-\frac{1}{\bar{z}_{\bar{O}}}\right| =\frac{1}{|z_O|^2}|z_O-z_{\bar{O}}|=\frac{2|\mathbf{Re}(z_O)|}{|z_O|^2},
\end{equation}
which in the limit $P_O\to \mathcal{I}^+_R$ and in Lorentzian signature means $\Delta =- 1/x_O^->0$. The result is then
\begin{equation}\label{lateretardedt} 
C_{\text{gen}}(\bar{B}\cup B) \approx 2 C_{\text{gen}}(B) -\frac{N}{3(\lambda x_O^-)^2}\left(2+\ln (\lambda x_O^-)^2+(\lambda x_O^-)^2\ln (\lambda x_O^-)^2\right),
\end{equation}
with $-\lambda x_O^-=e^{-\lambda(t_O-y_O)}$. The divergence of \eqref{lateretardedt} as $x_O^- \to 0$ (equivalently $P_O\to i^+_R$) should be understood as an artifact of extrapolating the \emph{eternal} RST saddle to arbitrarily late retarded times.
In our eternal setup the black hole parameter is held fixed (thermal equilibrium), so there is no net energy loss and the geometry does not capture the late-time backreaction associated with evaporation.
Therefore, the regime $x_O^- \to 0$ lies outside the domain of validity of the eternal solution. A faithful description of that limit requires working in the \emph{evaporating} RST background, where the late-time behavior is cut off by the time dependence of the geometry. In this work we will thus keep $x_O^-$ finite, and we do not interpret the growth of \eqref{lateretardedt} as a physical divergence of the capacity.

In comparison with the one interval case, it shares most of the general aspects from the previous case. The key difference is the time dependence induced by $\Delta$. This is not entirely surprising since a replica geometry with two fixed points is not $U(1)$--symmetric, and $\Delta$ represent a new energy scale in the configuration. Nevertheless it is interesting that the capacity can capture this while the entropy cannot.

The rapid growth in time that one finds in \eqref{lateretardedt} is a statement about the two--QES replica branch computed perturbatively near $n=1$.  The physical replica partition function is determined by the dominant branch of $-\log\Tr\rho^n$ at each $(n,t)$, and our calculation does not attempt to determine this
competition away from the $n\to1$ neighborhood.  From this point of view, a large two--QES capacity is
best interpreted as a diagnostic that higher $n$--derivatives probe global replica data (here
encoded in $\Delta$) more sharply than the entropy does, and that saddle selection near $n=1$ can be
delicate.  We comment on this mechanism and its relation to sharp features at the Page transition
in App. \ref{app:saddle_competition}.

\begin{figure}[t]
    \centering
    \includegraphics[width=9.8cm]{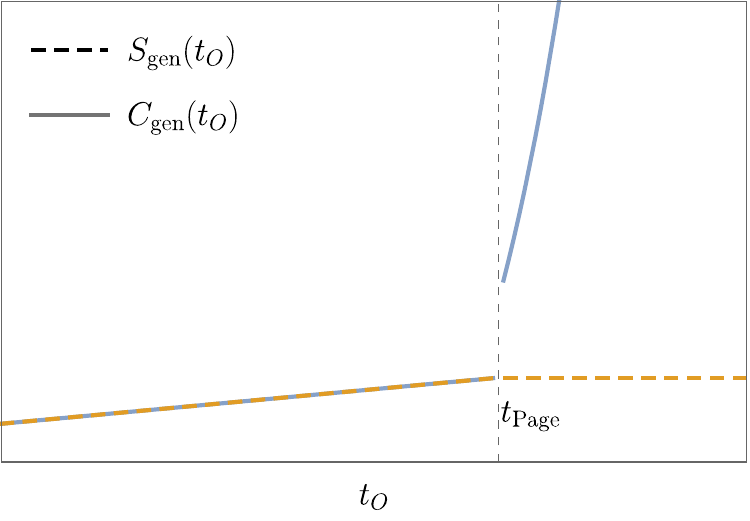}
    \caption{Qualitative comparison of generalized entropy and generalized capacity for two intervals as a function of $t_O$. Before $t_{\rm Page}$ both follow the same linear branch; after $t_{\rm Page}$ the entropy saturates while the capacity exhibits a jump and grows rapidly.}
    \label{fig: Cys}
\end{figure}

As illustrated in Fig.~\ref{fig: Cys}, the generalized entropy and the generalized
capacity display qualitatively different Page-time behavior in the two-interval setup.
For $t_O<t_{\rm Page}$ both quantities follow the no-island branch.
At $t_O=t_{\rm Page}$ the dominant saddle switches: $S_{\rm gen}$ saturates to the constant two-QES
plateau, while $C_{\rm gen}$ exhibits a discontinuous jump and then grows rapidly on the two-QES
branch.



\section{Discussion and outlook}
\label{conclu}

The main goal of this paper was to compute the capacity of entanglement in a model where the island
mechanism is under full analytic control in asymptotically flat space, namely in a setting where the specific conformal-welding obstruction familiar from JT+bath calculations is absent.  Working in the RST model
\cite{CGHS1992,RST1992,Polyakov1981}, we computed the gravitational contribution to the capacity in the eternal black hole background of \cite{HartmanShaghoulianStrominger2020}, both for a single
interval $\text{Rad}=[P_O,P_\Lambda]$ and for the two--interval configuration $\overline{\text{Rad}} \cup \text{Rad}$.  The technical lesson is simple. Since the capacity requires more details about the replica construction, it forces one to solve the replica equations \emph{globally}.  Local uniformization near the fixed point,
while convenient and physically transparent, does not by itself fix the homogeneous sector of the linearized solutions.  The constant mode is precisely what determines the on-shell variation that defines $C_{\rm grav}$, and it is fixed only after imposing global single-valuedness together with the appropriate fixed-state conditions at
infinity.  This global step is essentially absent from the entropy calculation. This is because to compute entanglement entropy, it is enough to have $\text{Tr} \rho_A^n$ at order $n-1$, which only uses the background solutions $\chi^{(1)}$ to compute areas. In contrast, the capacity of entanglement requires dependence at order $(n-1)^2$, and therefore the solutions of the replica equations at order $n-1$.

\subsection*{One interval}
For a single interval $\text{Rad}$ the final answer parallels the generalized entropy in the eternal black hole: the QES configuration always dominates, the results are constant in time and can be expressed in a compact form in terms of $P_O$. It is time independent as the replica construction for one interval preserves the $U(1)$ symmetry of the background eternal black hole geometry in the $P_O\to\mathcal{I}^+_R$ limit. The main difference with respect to $S_{\text{gen}}$ is that the capacity ended up UV divergence free. On the other hand,  it seems to be a common feature for both RST and JT gravity \cite{KawabataEtAl2105}  that $S_{\text{gen}}$ and $C_{\text{gen}}$ have the same constant gravitational contribution, the area of the QES.

\subsection*{Two intervals}

For two intervals $\overline{\text{Rad}}\cup \text{Rad}$, following the generalized entropy, the capacity is divided in two different regimes. First, the early-time regime (or no-island regime, where the Hawking saddle dominates) and then the late-time regime (or island regime, where the wormhole saddle dominates). For the latter, we computed $C_{\text{gen}}$ in the factorization regime and it showed a qualitative difference with respect to $S_{\text{gen}}$. The main point is that it is time dependent, due to a new energy scale appearing in the gravitational capacity $\Delta = |z_1-z_2| \sim -1/x_O^-$  (in Lorentzian signature). This scale represents the distance between the two quantum extremal surfaces, or the volume of the island that covered the black hole. A comparison graph between the entropy and the capacity is shown in Fig. \ref{fig: Cys}. We argued that although $C_{\text{gen}}$ becomes infinite as $P_O$ approaches $i^+$ through $\mathcal{I}^+_R$, the validity of our analysis breaks down if we accounted for the mass loss in the black hole due to radiation.

\subsection*{Topological phase transitions}
One of the motivations to study the capacity was the close parallel between thermodynamic and Rényi quantities (see table \ref{tab:analogy}). Heat capacity in thermodynamic systems can serve as a diagnostic tool for phase transitions (for example the magnetic susceptibility in the 2D Ising model), and so the CoE is an obvious quantity to probe the so called topological phase transition that gives place to the Page curve of the radiation. This analogy is made explicit as the modular definition of the capacity is given by \cite{deBoerJarvelaKeskiVakkuri2019}
\begin{equation}
    C(\rho) = \langle K^2\rangle - \langle K\rangle^2
\end{equation} 
with $K$ de modular Hamiltonian associated to he reduced density matrix $\rho_{\overline{\text{Rad}}\cup\text{Rad}}$. Whether the Page curve represents a phase transition could be diagnosed by $C_{\text{gen}}$ being discontinuous or divergent around the Page time. In Fig. \ref{fig: Cys} we showed that after the Page time the capacity loses the UV divergence given by $\ln \epsilon$, and although this computation is only valid after the factorization regime, we expect this UV constant term to be subtracted even in the vicinity of $t_{\text{page}}$. If $C_{\text{gen}}$ represents, loosely speaking, the spread of the eigenvalues in $\rho_{\overline{\text{Rad}}\cup\text{Rad}}$ then its divergent late-time behavior could be related to the extremely chaotic unitary evolution of the black hole's degrees of freedom.

\subsection*{Page times}

A useful caveat is that saddle dominance is defined at the level of the replicated partition
function \(Z_n(t\equiv t_O)\) (equivalently the on-shell action \(I_n(t)\)), so it is intrinsically a question
in the \((n,t)\) plane. The usual Page time $t_{\text{Page}}$ is the crossing at \(n=1\), but there is no general
reason for the transition curve \(t_{\text{Page}}(n)\) to be \(n\)-independent. Since the capacity probes
a second \(n\)-derivative at \(n=1\), it is especially sensitive to the local shape of this phase
boundary and can therefore develop a parametrically sharper feature even when \(S_{\rm gen}(t)\)
itself remains continuous. We comment on this mechanism in Sec.~\ref{subsec:two-interval_capacity} and give a more systematic discussion in App. \ref{app:saddle_competition}.

There are several obvious directions to push these results further.  First, our two--interval
computation relied on the factorization channel for the twist four-point function.  Beyond this
regime, the matter source is controlled by the full $\mathcal F_n(\eta,\bar\eta)$ of the twist
correlator \cite{CalabreseCardyTonni2009}, and the accessory parameters entering the constraint
equation become nontrivial functions of the cross-ratio.  From the QFT side, a systematic approach
is to treat $\log\mathcal F_n$ by OPE/Virasoro-block expansions and compute the resulting corrections
to the accessory parameters; in large-$c$ CFTs this structure is sharpened by semiclassical block
dominance \cite{Hartman2013}.  From the gravitational side, the same problem is naturally phrased as
a deformation of the meromorphic stress tensor source and the associated global monodromy
conditions.  It would be interesting to see whether the time-dependent terms we found persist once
these non-factorized contributions are included, and whether the competition of saddles in the
$(n,t)$ plane becomes smoother or sharper.

Second, the natural target is evaporation.  The RST model allows one to treat evaporating black hole
backgrounds analytically, and the island Page curve is already under excellent control in this
setting \cite{HartmanShaghoulianStrominger2020,GautasonEtAl2020}.  The capacity is a particularly
sharp observable in an evaporating geometry: unlike the eternal case, one expects a genuine
time-dependent competition between saddles even for the von Neumann entropy, and the non-uniformity
in $n$ that is only latent in the eternal problem may become dynamically important.  A controlled
computation of $C_{\rm gen}(t)$ in the evaporating RST background would provide a flat-space analog
of the JT analyses \cite{KawabataEtAl2102,KawabataEtAl2105} and would help clarify which features of
the capacity are universal and which are model-dependent.

Third, our results suggest that the capacity is a useful diagnostic of ``global replica data'' that
does not appear in $S_{\rm gen}$.  It would be worthwhile to connect this more directly to modular
observables in QFT.  In ordinary QFT, the capacity is the variance of the modular Hamiltonian and
can be interpreted as a measure of fluctuations of the entanglement spectrum
\cite{deBoerJarvelaKeskiVakkuri2019,Nandy2021}.  In holographic theories, modular data is related to
bulk modular energy and relative entropy \cite{JLMS2016}.  The RST setting provides a tractable
laboratory to explore which aspects of that relation survive in asymptotically flat, non-AdS
gravity, and to what extent the global replica solution we constructed can be rephrased in
intrinsically modular terms.

Finally, there is the more basic question of going beyond the perturbative neighborhood of $n=1$.
For the entropy this is often bypassed by geometric arguments (cosmic branes, refined Rényi
entropies) \cite{Dong2016}, but the RST model may allow a more explicit treatment: one can ask for
replica solutions at finite integer $n$ and study how the global constraints and state-fixing
conditions generalize.  Even partial control of this problem would sharpen the discussion of the
$(n,t)$ phase boundary and place the non-uniformity we found on firmer footing.  Whether the
time-dependent two--QES capacity persists as a property of the full integer-$n$ family, or is
reorganized by additional saddles at finite $n$, is a sharp question that RST seems uniquely well
suited to address.

We close with a broad takeaway.  The island formula makes the Page curve accessible in a wide class
of models, but it does not exhaust the replica physics.  The capacity of entanglement is the
simplest observable that forces one to keep track of global replica data and of saddle dominance in
a neighborhood of $n=1$.  In RST this can be done analytically, and the result shows that even in a
situation where the entropy has plateaued, higher modular cumulants can display qualitatively new
behavior.  Whether this is best viewed as a refinement of the Page-curve story or as a warning about
non-uniform analytic continuation depends on what ultimately controls the finite-$n$ replica
family.  Either way, the capacity appears to be an efficient probe of precisely the physics that is
otherwise hidden behind the smoothness of $S_{\rm gen}$.

\subsection*{Acknowledgments}
We would like to thank to Pedro Martinez and Guille Silva for useful discussions. DF is supported by fellowships from CONICET (Argentina).


\appendix

\section{The $U(1)$--symmetric ansatz}
\label{app:U1-ansatz}

In this appendix we present an ansatz for the replica solution in the $U(1)$--symmetric case for one interval, with a somewhat different interpretation for the global charge fixing in the solutions. 

Since we are dealing with one interval, the topology of the replicated space $\widetilde{\mathcal{M}}_n$ is the same as for $n=1$. This means that the general solution can be made explicit in terms of the solution for the background $n=1$ \eqref{eternal_BH_sol}, now depending on $\tilde{z}$
\begin{equation}
    \Omega = \frac{\widetilde{M}}{\tilde{\lambda}} +\tilde{\lambda}^2 \tilde{z}\bar{\tilde{z}}\,.
\end{equation}
Going to the orbifold changes $\tilde{\beta}=2\pi/\tilde{\lambda}\to \tilde{\beta}/n $, so we substitute $\tilde{\lambda}\to n\tilde{\lambda}$ in the solution. Using the uniformization map  
\begin{equation}
\tilde{\lambda}\tilde{z}=(\lambda(z-z_Q))^{1/n},    
\end{equation}
the proposed ansatz is
\begin{equation}\label{eq:U1-ansatz}
    \Omega^{(n)}(z,\bar{z}) = \frac{M}{n\lambda}+n^2\lambda^{2/n}|z-z_Q|^{2/n}\,.
\end{equation}

The key point is that we imposed $\widetilde{M}/\tilde{\lambda}\to M/\lambda$  under the map, since temperature is not a free parameter in RST gravity and the canonical partition function $Z(\tilde{\beta})$ is not going to be well defined (meaning not computable in any sensible way) for generic $\tilde{\beta}$, mainly because there are no smooth classical solutions for generic boundary conditions. Only for $\tilde{\beta} = \beta_{BH}$ we can perform the saddle point approximation and use the smooth eternal black hole solution.

Expanding around $n=1$ yields $\Omega = \Omega^{(1)}+(n-1)\delta\Omega$ with
\begin{equation}
    \delta\Omega = -\frac{M}{\lambda} +2\lambda^2 |z-z_Q|^2-\lambda^2|z-z_Q|^2\ln \lambda^2|z-z_Q|^2\,.
\end{equation}
To obtain the capacity one needs $\delta\chi$. We recall that $\chi$ is not a scalar, it transforms as
\begin{equation}
    \chi(x)\xrightarrow[x\to x(w)]{}\chi(x) +\frac{1}{2}\ln x'\bar{x}'\,.
\end{equation}
so we have
\begin{equation}
    \delta\chi = \delta\Omega -\frac{1}{2}\ln \lambda^2|z-z_Q|^2\,.
\end{equation}
The non-homogeneous transformation law, together with the singular nature of the uniformization map is what ruins the existence of the conserved current at order $n-1$. The capacity follows as
\begin{align}
\begin{split}
    C_{\text{gen}} & = \frac{NM}{6\lambda} +\frac{N}{2}\lambda y_O,
\end{split}
\end{align}
which is the result reported in the main text.

\subsection*{Solving the equations for generic $n$}
Here we investigate the generic $n$ solution and its global charges in the orbifold for the simple case of one interval. The first two orbifold equations of motion \eqref{orbifold_eqs}, for generic $n$ and $\lambda^2=1$, can be written as
\begin{align}
\begin{split}
    \partial\bar{\partial}(\Omega-\chi) &= \frac{\pi}{2}(1-1/n)\delta\,,\\
    \partial\bar{\partial}\Omega&= e^{2(\chi-\Omega)}.
\end{split}
\end{align}
The first reads $\chi-\Omega = -\frac{1}{2}(1-1/n)\ln |z-z_Q|^2$, and the second is then
\begin{equation}
    \partial\bar{\partial}\Omega = e^{2\frac{-1}{2}(1-1/n)\ln|z-z_Q|^2} = |z-z_Q|^{2(-1+1/n)},    
\end{equation}
which is solved by
\begin{equation}\label{omega_global_nfinito}
    \Omega = h(z)+g(z)+ n^2|z-z_Q|^{2/n}.
\end{equation}
The constraint equations is $\partial\Omega\partial\Omega-\partial\chi\partial\chi +\partial^2\chi = -\frac{6}{N}T_{zz}$ which gives
\begin{align}
\begin{split}
    \left(\partial\Omega\right)^2- \left[\partial\Omega-\frac{(1-1/n)/2}{z-z_Q}\right]^2+\partial^2\Omega +\frac{(1-1/n)/2}{(z-z_Q)^2}&=-\frac{6}{N}T_{zz},\\
 \partial^2\Omega +\frac{\partial\Omega (1-1/n)}{z-z_Q}+\frac{\frac{(1-1/n)}{2}\frac{(1+1/n)}{2}}{(z-z_Q)^2} &=-\frac{6}{N}T_{zz},\\
\end{split}
\end{align}
Plugin in \eqref{omega_global_nfinito} we get
\begin{equation}
    h''+\frac{\frac{(1-1/n)}{2}\frac{(1+1/n)}{2}}{(z-z_Q)^2} =-\frac{6}{N}T_{zz}.
\end{equation}
The stress energy tensor in the orbifold is given by its transformation rule (eq (23.5) of \cite{cardy-calabrese} taking $|v|=|z_O|\to\infty$, $w=z$ and $u=z_Q$ for single orbifold field) 
\begin{equation}
    -\frac{6}{N} T_{zz} = \frac{(n^2-1)/4n^2}{(z-z_Q)^2}.
\end{equation}
This gives $h''=0$ and therefore (restoring $\lambda^2$)
\begin{equation}
    \Omega = a+b(z-z_Q)+ c\overline{z-z_Q}+n^2\lambda^{2/n}|z-z_Q|^{2/n}.
\end{equation}
In dilatonic gravity the interdependent constants $a,b$ and $c$ are related to the total energy of the geometry. We finally remark that \eqref{eq:U1-ansatz} satisfies this with $a = M/n\lambda$ \footnote{It remains to be shown, in terms of a Komar-type of computation, that the total energy in the replicated space is $n$ times bigger than in the orbifold geometry ($\tilde{a}=M/\lambda=na$).} and $b=c=0$.

\section{Deriving $C_{\text{matter}}^{(n)}$ for an $n$ dependent conformal factor}\label{Cnmatter}
We go through basic computations around the replica trick in 2D conformally flat space and add the $n$ dependent conformal factor in the derivation, which is needed when the space is not fixed. This is somewhat analogous to the welding maps inducing contributions to the entanglement entropy in JT gravity.

The entanglement entropy is the $n\to1$ limit of the Rényi entropy
\begin{align}
\begin{split}
    S_A = \lim_{n\to1} S^{(n)}_A.
\end{split}
\end{align}
For a fixed flat metric we have
\begin{equation}
    ds^2_\mathbb{C}= dzd\bar{z},
\end{equation}
and the interval under consideration, $A$, will be $A= [z_1,z_2]$. Using the path integral representation of the density matrix it is known that
\begin{equation}
    Tr[\rho ^n ]_\mathbb{C}= \frac{Z[\widetilde{\mathcal{M}}_n]}{Z^n} = \langle \tau_n(z_1,\bar{z}_1)\tilde{\tau}_n(z_2,\bar{z_2})\rangle_{\mathcal{L}^{(n)}\mathbb{C}}.
\end{equation}
Where in the second equality we are already in the orbifold, which in this context is always $\widetilde{\mathcal{M}}/\mathbb{Z}_n = \mathbb{C}$, and $\tau_n$ are twist operator insertions. The replicated space has conical excesses around $\tilde{z}_{1,2}$. The uniformization map $\tilde{z}\to \tilde{z}(z)$ removes the conical excess and the space is smooth everywhere. When the space (fixed) is conformally flat we have
\begin{equation}
    ds^2 = e^{2\rho(z,\bar{z})}dzd\bar{z}
\end{equation}
We name the conformal transformation that takes us to flat space $z\to w(z)$
\begin{equation}
    ds^2_\mathbb{C} = dwd\bar{w} \to w'(z)\bar{w}'(\bar{z})dzd\bar{z} = e^{2\rho(z,\bar{z})}dzd\bar{z}=ds^2_g.
\end{equation}
Equation (17) from \cite{cardy-calabrese} relates ($\tau$ being primary fields of conformal weight $d_n$)
\begin{equation}\label{cardy-calabrese}
\langle \tau_n(z_1,\bar{z}_1)\tilde{\tau}_n(z_2,\bar{z}_2)\rangle_{\mathcal{L}^{(n)}\mathbb{C}}= |w'_1w'_2|^{d_n}\langle \tau_n(w_1,\bar{w}_1)\tilde{\tau}_n(w_2,\bar{w}_2)\rangle_{\mathcal{L}^{(n)}g}. 
\end{equation}
Also, since $|w'_1|= |e^{\rho_1}|$ we have 
\begin{equation}
    Tr[\rho^n]_g(w_1,w_2) = |e^{\rho_1+\rho_2}|^{-d_n}   Tr[\rho^n]_\mathbb{C}(z_1,z_2),
\end{equation}
where $w_j = w(z_j)$, $w'_j = w'(z_j)$ y $\rho_j = \rho(z_j,\bar{z}_j)$. From here it is straightforward that
\begin{equation}
    S_g^{(n)}[z_1,z_2] = S^{(n)}_\mathbb{C}[z_1,z_2] -\frac{d_n}{1-n}(\rho_1+ \rho_2)\qquad d_n = \frac{c}{12}\left(n-\frac{1}{n}\right).
\end{equation}
The CoE of matter is then
\begin{align}
    \begin{split}
        C_g^{(n)}= n^2\partial_n^2 \ln \text{Tr}\rho^n_g = n^2\partial^2_n \ln \text{Tr}\rho^n_\mathbb{C} - n^2\partial^2_n d_n (\rho_1+\rho_2)
        = C_\mathbb{C}^{(n)} +\frac{c}{6n} (\rho_1+\rho_2),
    \end{split}
\end{align}
giving
\begin{equation}
C_g^{\text{matter}}[z_1,z_2] = C_\mathbb{C}^{\text{matter}}[z_1,z_2] + \frac{c}{6}(\rho(z_1,\bar{z}_1)+\rho(z_2,\bar{z}_2)).
\end{equation}
The conformal factor in this construction, specially in \eqref{cardy-calabrese}, is the one from the orbifold. What is happening is that in flat or conformally flat spaces, this conformal factor coincides with the background one.

If we are interested in the replica trick in gravity, the conformal factor in the orbifold is dynamic, it must be obtained from equations of motion in the saddle point approximation. Also, and most distinctively, in gravity $\widetilde{\mathcal{M}}_n/\mathbb{Z}_n\neq \mathbb{C}$, in fact $\widetilde{\mathcal{M}}_n/\mathbb{Z}_n =\mathcal{M}_n$ (which generically has conical singularities in co-dimension 2 surfaces). This detail amounts to consider $\rho=\rho^{(n)}$ and caring about derivatives. The Rényi entropy gives
\begin{equation}
    S_g^{(n)}[z_1,z_2] = S^{(n)}_\mathbb{C}[z_1,z_2] -\frac{d_n}{1-n}(\rho^{(n)}_1+ \rho^{(n)}_2),\qquad d_n = \frac{c}{12}\left(n-\frac{1}{n}\right).
\end{equation}
When expanding $\rho^{(n)}=\rho + (n-1)\delta\rho$ it is clear that EE does not see $\delta\rho$, since $\frac{d_n}{1-n} = \frac{c}{12}\frac{n+1}{n}$ and when $n\to1$, terms involving $\delta\rho$ vanish. Instead, CoE does capture some details from the correction 
\begin{align}
    \begin{split}
        C_g^{(n)}&= n^2\partial_n^2 \ln \text{Tr}\rho^n_g = n^2\partial^2_n \ln \text{Tr}\rho^n_\mathbb{C} - n^2\partial^2_n d_n (\rho^{(n)}_1+\rho^{(n)}_2)\\
        &= C_\mathbb{C}^{(n)} -n^2\left[\partial^2_n d_n \rho^{(n)}_{1+2}+2\partial_n d_n \partial_n \rho^{(n)}_{1+2}+d_n \partial^2_n \rho^{(n)}_{1+2}\right]\\
        &= C_\mathbb{C}^{(n)} +\frac{c}{6n} (\rho_1+\rho_2) -\frac{c}{6}(n^2+1)(\delta\rho_1+\delta\rho_2)+ d_n (\delta^2\rho_1+\delta^2 \rho_2).
    \end{split}
\end{align}
Thus the capacity of entanglement of the matter sector for an interval $A= [z_1,z_2]$ gives
\begin{equation}
C_g^{\text{matter}}[z_1,z_2] = C_\mathbb{C}^{\text{matter}}[z_1,z_2] + \frac{c}{6}(\rho(z_1,\bar{z}_1)+\rho(z_2,\bar{z}_2)) - \frac{c}{3}(\delta\rho_1+ \delta\rho_2).
\end{equation}

\section{Matter stress tensor and the origin of the time dependence in the two--QES capacity}\label{app Tmatter}

In the two--interval configuration $\bar B\cup B$, the replica  saddle contains two fixed
points $\Sigma=\{z_1,z_2\}$ and the linearized orbifold equations are sourced by the matter stress
tensor in the presence of twist insertions.  The purpose of this appendix is twofold.  First, we
re-derive the form of the meromorphic source $\delta T^{\rm mat}_{zz}$ used in the main text, and
explain precisely what is assumed in the ``factorization'' regime.  Second, we show that the
time-dependent contribution to the two--QES capacity arises from a geometric invariant
$\Delta\equiv|z_1-z_2|$ and therefore persists under the controlled corrections to $\delta T_{zz}$
that follow from the non-factorized four--point function of twists.  We conclude with a short
discussion of why a time-dependent capacity is physically viable even when the generalized
entropy has entered its Page plateau.

On the orbifold, the matter sector is a CFT with $c=N$ coupled to the classical
background determined by the RST fields.  In the replica computation for two intervals, the matter
path integral reduces to a four--point function of twist operators,
\begin{equation}
G_n(z_1,z_2,z_O,\bar z_O)\equiv
\langle \tau_n(z_1)\,\tilde\tau_n(z_O)\,\tau_n(z_2)\,\tilde\tau_n(z_{\bar{O}})\rangle,
\end{equation}
and the holomorphic Ward identity fixes the one--point function of the stress tensor in this
background to be a meromorphic function with at most double poles at the insertion points:
\begin{equation}
\langle T(z)\rangle_{G_n}
=
\sum_{p\in\{z_1,z_2,z_O,\bar z_O\}}
\left(\frac{h_n}{(z-p)^2}+\frac{1}{z-p}\,\partial_p\ln G_n\right),
\qquad
h_n=\frac{c}{24}\Big(n-\frac1n\Big).
\label{app:Ward}
\end{equation}

The coefficients of the simple poles,
\begin{equation}
A_p(n)\equiv \partial_p\ln G_n,
\end{equation}
are the (holomorphic) accessory parameters.  In our orbifold equations only the singularities at
the fixed points $z_1,z_2$ enter the local source term when working in the limit $|z_{O,\bar{O}}|\to\infty$.  Expanding \eqref{app:Ward} near $n=1+\varepsilon$ gives $h_n=\frac{c}{12}\varepsilon+O(\varepsilon^2)$,
and it is convenient to isolate the order-$\varepsilon$ meromorphic source at the fixed points in
the normalized form used in the main text,
\begin{equation}
-\frac{6}{N}\,\delta T^{\rm mat}_{zz}(z)
=
\sum_{i=1}^2
\left[
\frac{1}{2}\frac{1}{(z-z_i)^2}
-\frac{A_i}{z-z_i}
\right],
\qquad
A_i\equiv \frac{6}{N}\,\partial_{z_i}S^{\rm flat}_{\rm vN},
\label{app:deltaTform}
\end{equation}
where $S^{\rm flat}_{\rm vN}$ is the CFT von Neumann entropy for the appropriate interval, $\bar{B}\cup B$, in the
orbifold coordinates (so that $A_i$ is the $n\to1$ limit of the corresponding accessory parameter at
the fixed point).  

In the factorization regime used in Sec.~3.2, the four--point function of twists is dominated by a
single OPE channel and takes the approximate form 
\begin{equation}
G_n(z_1,z_2,z_O,z_{\bar{O}})
\;\simeq\;
\langle \tau_n(z_1)\tilde\tau_n(z_O)\rangle\;
\langle \tau_n(z_2)\tilde\tau_n(z_{\bar{O}})\rangle.
\label{app:factorization}
\end{equation}
Since the two--point function of twists is fixed by conformal symmetry, this implies
\begin{equation}
S^{\rm flat}_{\rm vN}[z_1,z_O]\propto \frac{N}{6}\ln|z_1-z_O|^2,
\qquad
S^{\rm flat}_{\rm vN}[z_2,z_{\bar{O}}]\propto \frac{N}{6}\ln|z_2-z_{\bar{O}}|^2,
\end{equation}
and therefore the leading accessory parameters in \eqref{app:deltaTform} are
\begin{equation}
A_1=\frac{1}{z_1-z_O},
\qquad
A_2=\frac{1}{z_2-z_{\bar{O}}}.
\label{app:Ai-fact}
\end{equation}

Beyond the approximation \eqref{app:factorization}, conformal symmetry implies the standard
decomposition
\begin{equation}
G_n
=
(\text{fixed prefactor})\times \mathcal{F}_n(\eta,\bar\eta),
\qquad
\eta\equiv \frac{(z_1-z_O)(z_2-z_{\bar{O}})}{(z_1-z_2)(z_O- z_{\bar{O}})},
\label{app:crossratio}
\end{equation}
where $\eta$ is the cross ratio for the four insertion points.  The nontrivial function
$\mathcal{F}_n(\eta,\bar\eta)$ encodes the exchange of non-identity operators in the chosen channel
and it is precisely this quantity that is set to~$1$ in the strict factorization limit.  Since
$\tau_{n=1}$ is the identity, one has $\mathcal{F}_{n=1}(\eta,\bar\eta)=1$ and therefore
\begin{equation}
\ln \mathcal{F}_n(\eta,\bar\eta)
=
(n-1)\,f_1(\eta,\bar\eta)
+\frac12 (n-1)^2 f_2(\eta,\bar\eta)
+O\!\big((n-1)^3\big).
\label{app:Fexp}
\end{equation}
The correction to the accessory parameters at $n\to1$ is then
\begin{equation}
\delta A_i \equiv \lim_{n\to1}\frac{1}{n-1}\partial_{z_i}\ln\mathcal{F}_n
=\partial_{z_i} f_1(\eta,\bar\eta),
\label{app:dAi}
\end{equation}
so the exact stress tensor differs from \eqref{app:deltaTform} by the addition of these
$\delta A_i$ to \eqref{app:Ai-fact}.  In the OPE regime $|\eta|\ll 1$ in the same channel, the
standard conformal block expansion implies
\begin{equation}
\mathcal{F}_n(\eta,\bar\eta)=1+O\!\left(|\eta|^{\Delta_{\rm gap}}\right),
\qquad
\Delta_{\rm gap}>0,
\label{app:OPE}
\end{equation}
and hence $\delta A_i=O(|\eta|^{\Delta_{\rm gap}})$.  In Lorentzian kinematics, the relevant channel
for the late-time factorization corresponds to $|\eta(t_O)|\to 0$ exponentially fast; in particular,
for the Hartman two-sided configuration one finds 
\begin{equation}
|\eta(t_O)| \sim \exp\{-2\lambda(|t_O|-y_O)\}\qquad (|t_O|\gg y_O),
\label{app:eta-scaling}
\end{equation}
so the corrections $\delta A_i$ are exponentially suppressed at late times.  Since the time
dependence in our final expression for the capacity originates from a \emph{leading} geometric
invariant, the suppressed corrections \eqref{app:dAi} cannot remove it by any smooth
deformation: they merely produce subleading corrections to the QES locations and to the capacity.
Equivalently, within the regime of validity of the factorized channel, the time dependence is
robust under the controlled inclusion of $\mathcal{F}_n(\eta,\bar\eta)$.

A potential subtlety is the ordering of limits: the factorization assumption concerns the twist four-point function at fixed replica index $n$, whereas the capacity requires taking $n$-derivatives at $n=1$. In principle these operations need not commute, since $\log \mathcal F_n(\eta,\bar\eta)$ can contribute at order $(n-1)^2$ even when its leading effect on the von Neumann entropy is suppressed. 
This kind of subtlety is naturally phrased in the replica--defect language: as \(n\to 1\), defect OPE data can generate contact-term contributions controlled by operators that become the stress tensor in the limit, so the order of limits (analytic continuation in \(n\) versus kinematic/OPE limits) can matter in principle~\cite{Balakrishnan:2019replica}. In the factorization channel $|\eta|\ll 1$ relevant for our analysis, however, the OPE implies $\mathcal F_n=1+O(|\eta|^{\Delta_{\rm gap}})$ and therefore any such non-commutativity affects $A_i$ and $C_{\rm mat}$ only at $O(|\eta|^{\Delta_{\rm gap}})$, i.e. parametrically subleading in the same regime where factorization is valid.

\section{Saddle competition in the $(n,t)$ plane and the determination of the relevant transition time for the capacity}
\label{app:saddle_competition}

Before turning to the detailed derivation, it is worth stressing that the logic behind this
appendix is a direct gravitational analogue of a standard mechanism in statistical mechanics.
Rényi quantities naturally define an effective ``free energy'' as a function of the replica index
\(n\), and in a variety of QFT settings \(n\) can be reinterpreted as an inverse temperature for an
auxiliary thermal problem (most explicitly for spherical entangling surfaces, where the vacuum is
conformally mapped to a thermal state on \(\mathbb{R}\times\mathbb{H}^{d-1}\))~\cite{Casini:2011kv}.
Once multiple saddles (or phases) are available, the replicated partition function is controlled by
the dominant branch, and non-analyticities in \(n\)-derivatives arise in the same way that heat
capacity singularities arise from branch switching in ordinary thermodynamics. This phenomenon has
been explored extensively in holography, where varying \(n\) changes the temperature of the dual
hyperbolic black hole and can trigger genuine phase transitions in Rényi entropies~\cite{Belin:2013dva},
and it also appears already in the classic two-interval problem as a change of dominant channel
visible in Rényi data~\cite{Headrick:2010zt}. More recently, it has become clear that the
neighborhood of \(n=1\) can be especially subtle in gravitational settings with competing extremal
surfaces, because the correct continuation may depend on the order of optimization away from
\(n=1\)~\cite{Dong:2023mcb}. In the specific context of islands and replica wormholes, the same
general moral has been emphasized for the capacity of entanglement: while the generalized entropy
can remain continuous at the Page transition, the capacity can exhibit a sharper feature precisely
because it probes higher derivatives with respect to \(n\) and hence ``feels'' the competition
problem in the \((n,t)\) plane~\cite{KawabataEtAl2105}. This appendix makes this structure explicit in our RST computation and clarifies which transition curve controls the
features of the capacity.

The semiclassical replica computation determines $\Tr(\rho_A^n)$ from a sum over replica saddles.
In the semiclassical limit this reduces to a competition between on-shell actions,
\begin{equation}
\Tr(\rho_A^n)\;\approx\;\sum_{a}\exp\!\big[-I_n^{(a)}(t)\big],
\qquad
I_n(t)\equiv -\log\Tr(\rho_A^n)=\min_a I_n^{(a)}(t),
\end{equation}
where $a$ labels candidate saddles. In the two-sided two-interval problem $A=\overline{\text{Rad}}\cup \text{Rad}$ (the endpoints of these regions induces a "time dependence" in the partition function $t\equiv t_O$) the relevant families are the no-island saddle ($a=0$) and the two-QES island saddle ($a=2$). The transition line in the $(n,t)$ plane is defined by
\begin{equation}
I_n^{(0)}\big(t_*(n)\big)=I_n^{(2)}\big(t_*(n)\big).
\label{eq:transition_line_def}
\end{equation}
It is useful to introduce the replica ``free energy''
\begin{equation}
F(n,t)\equiv \frac{1}{1-n}\log \Tr(\rho_A^n)=\frac{I_n(t)}{n-1},
\end{equation}
so that \eqref{eq:transition_line_def} is equivalent to $F^{(0)}(n,t_*)=F^{(2)}(n,t_*)$ for $n>1$.

Near $n=1$ each saddle admits an expansion
\begin{equation}
\frac{I_n^{(a)}(t)}{n}= I_1(t) + (n-1)\,S^{(a)}(t) + \frac12 (n-1)^2\,C^{(a)}(t) + O\!\big((n-1)^3\big),
\label{eq:In_expand}
\end{equation}
where $S^{(a)}(t)$ is the (generalized) von Neumann entropy evaluated on saddle $a$, and $C^{(a)}(t)$ is the corresponding capacity coefficient. Since the term $I_1(t)$ is common to all saddles, saddle dominance near $n=1$ is controlled by the reduced free energy
\begin{equation}
\widetilde F^{(a)}(n,t)\equiv S^{(a)}(t)+\frac12(n-1)\,C^{(a)}(t)+O\!\big((n-1)^2\big),
\end{equation}
and the transition condition becomes
\begin{equation}
\Delta S(t)+\frac{n-1}{2}\,\Delta C(t)=0,
\qquad
\Delta S\equiv S^{(0)}-S^{(2)},\qquad
\Delta C\equiv C^{(0)}-C^{(2)}.
\label{eq:transition_condition_linearized}
\end{equation}
The usual Page time for the von Neumann entropy is obtained by taking $n\to 1$ first, in which case \eqref{eq:transition_condition_linearized} reduces to $\Delta S(t_{\rm Page})=0$. However, for any fixed $n>1$ the transition is controlled by \eqref{eq:transition_condition_linearized} rather than by $\Delta S=0$, and the quadratic coefficient $\Delta C$ shifts the transition line away from the $n=1$ crossing.

In our setup $\Delta S(t)$ has the familiar behavior: in the no-island saddle it grows linearly at late times, while the island saddle is time-independent, so $\Delta S(t)$ crosses zero at the standard Page time $t_{\rm Page}$ determined from the entropy. The novel feature in the two-QES saddle is that the capacity coefficient $C^{(2)}(t)$ can be parametrically large once evaluated on the Lorentzian continuation of the island kinematics. In particular, inserting the Lorentzian separation of the two fixed points into the two-QES result yields a rapidly growing $C^{(2)}(t)$ once the invariant $\Delta(t)=|z_1-z_2|$ becomes large. Consequently $\Delta C(t)=C^{(0)}(t)-C^{(2)}(t)$ is large and negative for $t\gtrsim t_{\rm Page}$, and \eqref{eq:transition_condition_linearized} implies that the region in which the two-QES saddle dominates is confined to an extremely small wedge $1<n<n_*(t)$ above $n=1$:
\begin{equation}
n_*(t)-1 \;\simeq\; \frac{2\,\Delta S(t)}{|\Delta C(t)|}\qquad (t>t_{\rm Page}).
\label{eq:ncrit_def}
\end{equation}
This wedge shrinks rapidly as $t$ increases because $|\Delta C(t)|$ grows much faster than $\Delta S(t)$.

The physical consequence is that the ``Page time'' inferred from the von Neumann entropy (the $n\to1$ crossing) does not necessarily control the behavior of higher $n$-derivatives. In particular, while the entropy is governed by the linear term in \eqref{eq:In_expand}, the capacity is governed by the quadratic term and is therefore sensitive to how saddle dominance changes in a neighborhood of $n=1$. When $\Delta C(t)$ is large, the transition line $t_*(n)$ becomes extremely steep in the $(n,t)$ plane. Equivalently, the limits $n\to1$ and $t\to t_{\rm Page}$ fail to commute uniformly. This non-uniformity provides a natural mechanism for sharp features in the capacity (jumps or peaks) at or near the entropy Page transition, and it clarifies why the time scale controlling the onset of island dominance for $\Tr(\rho^n)$ at $n>1$ can be parametrically different from the von Neumann Page time.

The limited control near $n=1$ that our computation has is already enough to clarify the qualitative physics of the capacity at the Page transition. The von Neumann Page time is determined by the crossing at $n=1$, i.e.\ by $\Delta S(t_{\rm Page})=0$, and the generalized entropy can remain continuous across the transition. By contrast, the capacity is sensitive to the \emph{quadratic} coefficient in $(n-1)$, and therefore to how saddle dominance changes in a neighborhood of $n=1^+$. As a result, even if the entropy exhibits only a kink, the capacity may develop a sharp feature (a jump or a peak) at the von Neumann Page transition whenever $\Delta C(t_{\rm Page})\neq 0$. This provides a natural mechanism---fully consistent with the perturbative nature of our computation---by which higher cumulants of the entanglement spectrum can display nontrivial behavior at the Page transition, without requiring control of the Rényi saddles at finite integer $n$.

\paragraph{Related non-gravitational examples.}
The distinction between the contribution that dominates $\Tr\rho_A^n$ at integer $n>1$ and the analytic continuation that defines the von Neumann entropy at $n\to1$ is not unique to gravity; it is a generic feature of replica constructions whenever multiple channels (or phases) compete. A standard QFT example is the entanglement of two disjoint intervals in a 2D CFT, where $\Tr\rho_A^n$ is controlled by a four-point function of twist operators and depends on a nontrivial function of the cross-ratio, so that different OPE channels can dominate in different regimes \cite{CalabreseCardyTonni2009}. In the large-$c$/sparse-spectrum limit this channel dominance can be sharpened semiclassically in terms of Virasoro blocks, and the resulting structure for Rényi entropies versus the $n\to1$ continuation is discussed in detail in \cite{Hartman2013}. In condensed-matter settings, monitored random circuits and measurement-induced entanglement transitions provide another concrete arena where the effective replica description depends explicitly on $n$ and the continuation to $n\to1$ can be conceptually distinct from the physics at fixed integer Rényi index \cite{SkinnerRuhmanNahum2019}. These examples support the general lesson emphasized in this appendix: the transition time inferred from the $n\to1$ entropy need not coincide with the transition scale controlling Rényi observables at $n>1$, and higher $n$-derivatives (such as the capacity) can be especially sensitive to this non-uniformity.

\bibliographystyle{unsrturl}
\bibliography{references}

\end{document}